\documentclass[a4paper,10pt]{article}

\newcommand{\eps}{\varepsilon}
\newcommand{\rmd}{{\rm d}}

\usepackage[lmargin=2cm,rmargin=2cm,tmargin=2cm,bmargin=2cm]{geometry}
\usepackage{amssymb, amsmath, graphicx, url}
\DeclareMathOperator{\tr}{tr}

\title{Loschmidt Echo}

\author{Arseni~Goussev$^1$, Rodolfo~A.~Jalabert$^2$,
  Horacio~M.~Pastawski$^3$, and
  Diego~Wisniacki$^4$\\ \vspace{-0.3cm}\\ {\small $^1$Max Planck
    Institute for the Physics of Complex Systems, N{\"o}thnitzer
    Stra{\ss}e 38, D-01187 Dresden, Germany}\\ {\small $^2$Institut de
    Physique et Chimie des Mat{\'e}riaux de Strasbourg, UMR 7504,
    CNRS-UdS,}\\ {\small 23 rue du Loess, BP 43, 67034 Strasbourg
    Cedex 2, France}\\ {\small $^3$Instituto de F{\'i}sica Enrique
    Gaviola (CONICET-UNC) and Facultad de Matem{\'a}tica
    Astronom{\'i}a y F{\'i}sica,}\\ {\small Universidad Nacional de
    C{\'o}rdoba, Ciudad Universitaria, C{\'o}rdoba 5000,
    Argentina}\\ {\small $^4$Departamento de F{\'i}sica, FCEyN, UBA,
    Ciudad Universitaria, Buenos Aires C1428EGA Argentina}}

\date{\today}

\begin{document}

\maketitle

The Loschmidt echo is a measure of the revival occurring when an
imperfect time-reversal procedure is applied to a complex quantum
system. It allows to quantify the sensitivity of quantum evolution to
perturbations. An initial quantum state $| \psi_0 \rangle$ evolves
during a time $t$ under a Hamiltonian $H_1$ reaching the state $|
\psi_t \rangle$. Aiming to recover the initial state $| \psi_0
\rangle$ a new Hamiltonian $-H_2$ is applied between $t$ and
$2t$. Perfect recover of $| \psi_0 \rangle$ would be achieved by
choosing $H_2$ to be equal to $H_1$. This is not possible in realistic
setups, and there always appears a difference between $H_2$ and $H_1$,
leading to a non-perfect recovery of the initial state. The forward
evolution between $t$ and $2t$ under the Hamiltonian $-H_2$ is
equivalent to a backward evolution from $t$ to 0 under $H_2$, which
embodies the notion of time-reversal. The Loschmidt echo studies are
focused on the cases where the dynamics induced by the Hamiltonians
$H_1$ and $H_2$ are non-trivial, or sufficiently complex (like that of
a classically chaotic one-particle system or a many-body system).

\section{Introduction}

The concept of time-reversal has captured the imagination of
physicists for centuries, leading to numerous vivid discussions. An
emblematic example of these was the controversy around the second law
of thermodynamics between Ludwig Boltzmann and Joseph Loschmidt. When
Boltzmann was trying to develop the microscopic theory of the second
law of thermodynamics, Loschmidt raised an objection that had profound
influence on the subsequent development of the theory. He argued that,
due to the time-reversal invariance of classical mechanics, evolution
in which the entropy can decrease must exist. These states could be
reached by reversing velocities of all molecules of the system. After
such a reversal the entropy would then no longer grow but decrease,
seemingly violating the second law of thermodynamics. As a response
Boltzmann argued that such a time-reversal experiment would be
impossible and put forward a statistical interpretation of the second
law. Undoubtedly, Boltzmann's argument is a valid approach to the
problem of the ``arrow of time'' for generic macroscopic
systems. However, in quantum systems with few degrees of freedom,
today's technological advances make it meaningful to address
time-reversal experiments.

\subsection{Pioneering experiments}

{\bf Refs.~\cite{Hah50Spin, RPW71Time, BH84Atomic, ZME92Polarization,
    LUP98Attenuation}}

\medskip

\noindent The first controlled experimental implementation of
time-reversal was achieved in the fifties with the inversion of
nuclear spins precessing around a magnetic field. Such a reversal was
able to compensate the local-field inhomogeneities responsible for the
decay of the free induction signal. The time-reversal was implemented
through a radio-frequency pulse, leading to the formation of a revival
in the induction signal, known as spin echo or Hahn echo. Early on
Erwin Hahn recognized that his procedure, which he viewed as a change
in the sign of the system Hamiltonian, provided a quantum
implementation of the Loschmidt proposal. Interactions between spins,
not reversed in Hahn's procedure, were the main cause of the decay of
the echo, within a time scale known as $T_2$.  This experiment was
followed by a number of variants, both within magnetic resonance as
well in other time-dependent spectroscopies. In particular, the
dynamical decoupling techniques implemented in quantum registries to
isolate them from their environment are variants of Hahn's original
experiment.

The next level of complexity was developed with the reversal of the
many-body interactions. No general recipe is available in this case,
but the sign of the truncated spin-spin dipolar interaction, which is
associated with the cosine of the angle between the inter-spin vector
and the quantizing magnetic field, can be reversed by rotating the
spins into a quantization axis. This was the proposal of the Magic
Echo procedure, implemented by Won-Kyu Rhim, Alex Pines and John
Stewart Waugh in the seventies.
%, the so-called Warren pulse sequence.
Substantially simpler is the Polarization Echo sequence, implemented
by Richard Ernst and collaborators in the nineties. In this last case
the initial state has a local nature as it is labeled by the presence
of a rare $^{13}$C which plays the role of a local probe to inject and
later on detect the polarization of a nearby $^1$H immersed in a $^1$H
network.

This idea was further exploited by Patricia Levstein, Horacio
Pastawski and Gonzalo Usaj to test the stability of many-body
dynamics. They suggested that the inefficiency of the time-reversal
procedure found in all the previous experiments has a connection with
quantum chaos and the inherent dynamical complexity of the many-body
spin system.

\subsection{Definition}

{\bf Refs.~\cite{Per84Stability, JP01Environment}}

\medskip

\noindent The Loschmidt echo is defined as
\begin{equation}
M(t) = \left| \langle \psi_0 | e^{i H_2 t / \hbar} e^{-i H_1 t /
  \hbar} | \psi_0 \rangle \right|^2
\label{LE_definition}
\end{equation}
where
\begin{itemize}

\item $| \psi_0 \rangle$ is the state of the system at time 0

\item $H_1$ is the Hamiltonian governing the forward evolution

\item $H_2$ is the Hamiltonian governing the backward evolution

\item $t$ is the instant at which the reversal takes place

\end{itemize}

\begin{figure}[ht]
\center
\includegraphics[width=3.5in]{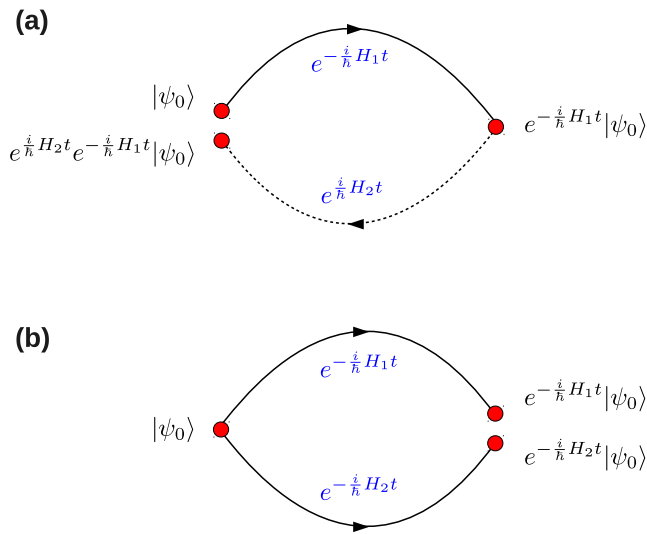}
\caption{Schematic flow of the time-evolution for (a) the Loschmidt
  echo and (b) the fidelity.}
\label{fig1}

\end{figure}

\begin{figure}[ht]
\center
\includegraphics[width=3in]{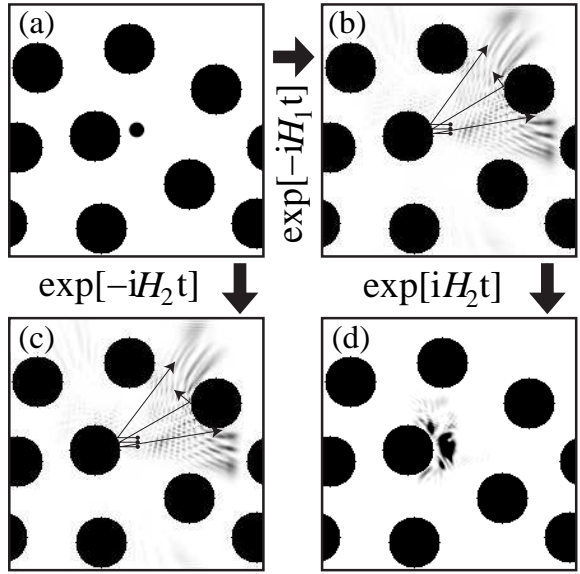}
\caption{Wave-packet evolution in a Lorentz gas. (a) Initial state at
  time $t=0$ with momentum pointing to the left. (b) State evolved
  with the Hamiltonian $H_1$ in the interval $(0,t)$. (c) State
  evolved with the Hamiltonian $H_2$ in the same time interval. (d)
  State evolved from that depicted in panel (b) with the $-H_2$ for
  the time interval $(t,2t)$. In panels (b) and (c) the classical
  trajectories corresponding to three initial positions within the
  original wave-packet are shown for reference. The square of the
  overlap between the states (a) and (d), the Loschmidt Echo, is
  $M(t)=0.09$, the same as that between the states of panels (b) and
  (c). Adapted from Ref.~\cite{CPW02Decoherence}. Copyright (2002),
  American Physical Society.}
\label{fig2}
\end{figure}

The time evolution appearing in Eq.~(\ref{LE_definition}) is
schematically represented in Fig.~\ref{fig1}(a), where the Loschmidt
echo quantifies the degree of irreversibility. Alternatively,
Eq.~(\ref{LE_definition}) can be interpreted as the overlap at time
$t$ of two states evolved from $| \psi_0 \rangle$ under the action of
the Hamiltonian operators $H_1$ and $H_2$. In this case the Loschmidt
echo is a measure of the sensitivity of quantum evolution to
perturbations. The quantity $M(t)$ interpreted in this manner is
illustrated in Fig.~\ref{fig1}(b) and is usually referred to as
fidelity.  The equivalence between Loschmidt echo and fidelity is
displayed in Fig.~\ref{fig2} for the example of an initially localized
wave-packet in a Lorentz gas.  The probability density of the evolved
state under $H_1$ ($H_2$) is plotted in Fig.~\ref{fig2} (b) (c), while
Fig.~\ref{fig2}(d) presents the state resulting from the combined
evolution $H_1$ and then $-H_2$ both for a time $t$. The overlap
squared between the states (a) and (d) defines the Loschmidt echo,
while the overlap squared between the states (b) and (c) defines the
fidelity. It is important to remark that even though the probability
density distributions in figures (b) and (c) are seemingly identical,
the phase randomization due to the difference between the two
Hamiltonians leads to a weak Loschmidt echo.

The definition given by Eq.~(\ref{LE_definition}) assumes that the
Hamiltonian operators $H_1$ and $H_2$ are independent of time. A
generalization to the case of time dependent Hamiltonians is
straightforward.  The case of non-Hermitian operators $H_1$ and $H_2$
can also be considered, but in such a case the equivalence between the
Loschmidt echo and fidelity would not hold.

\subsection{Loschmidt echo and decoherence}

{\bf Refs.~\cite{Zur01Sub, CDPZ03Decoherence, Zur03Decoherence,
    CPJ04Universality, CGW10Discrepancies, Bon11Lyapunov}}

\medskip

\noindent In an isolated quantum system the evolution is unitary;
initially pure states remain pure. When the system is connected to the
external world the purity of an initial state is generically lost in
the course of time evolution. This ubiquitous process is referred to
as decoherence. The coupling to the environment degrees of freedom, or
alternatively the lack of complete knowledge of the system
Hamiltonian, typically wash the specific interference properties that
would be observed within a unitary evolution. Since interference is
one of the most important signatures of quantum mechanics, decoherence
has both, fundamental and practical interest. Concerning the former
aspect, decoherence has been proposed as a road towards the classical
behavior observed in macroscopic systems. On the second aspect, it is
clear that decoherence represents a limitation in the implementation
of quantum computers or nanoscopic devices based on quantum effects,
specially when the number of qubits is scaled up, increasing the
complexity of the system. The Loschmidt echo, by including a
non-controlled part of the Hamiltonian, or the noisy effects of the
environmental degrees of freedom, gives a way to quantify decoherence
effect. By attempting the time-reversal of the controlled part of the
Hamiltonian, one can quantify at which rate the neighborhood of the
system (either a unitary part of larger system Hamiltonian or though
the effect of infinite uncontrolled environmental degrees of freedom)
acts as decoherence. Moreover, other quantifiers of decoherence, such
as for instance the purity, can be shown, under certain assumptions,
to be characterized by the same dependence on the underlying classical
dynamics as the Loschmidt echo.

\subsection{Broad interest of the Loschmidt echo}

Since the beginning of this century, the Loschmidt echo has been a
subject of intensive studies by researchers from different scientific
communities. A list of problems, in which the concept of the Loschmidt
echo appears naturally, includes
\begin{itemize}

\item Quantum chaos, or the quantum theory of classically chaotic systems

\item Decoherence or the emergence of ``classical world'' (open
  quantum systems)

\item Quantum computation and quantum information

\item Spin echo in Nuclear Magnetic Resonance

\item Linear waves (Elastic waves, microwaves, time-reversal mirrors,
  etc.)

\item Nonlinear waves (Bose-Einstein condensates, Loschmidt cooling, etc.)

\item Statistical mechanics of small systems

\item Quantum chemistry and molecular dynamics

\item Quantum phase transitions and quantum criticality

\item Mathematical aspects of the Loschmidt echo

\end{itemize}

Such broad interest stems from the fact that the Loschmidt echo is a
measurable quantity exhibiting, in certain regimes, a robust and
universal behavior. It is by means of the forward and backward
time-evolutions that the Loschmidt echo filters out some uninteresting
effects, while amplifying other, more important physical processes
taking place in complex quantum systems.

\section{Loschmidt echo: techniques and main results}

The Loschmidt echo is typically a decreasing function of $t$. It is
therefore of foremost importance to determine the form of the decay
and the associated characteristic times in various physical situations
as a function of the characteristic quantities of the problem. Even
though the Loschmidt echo has been addressed in a variety of
scenarios, the main effort has been directed towards the simplest
set-up of one-body dynamics. Well established results for the case of
one-body systems, together with techniques adapted from the field of
quantum chaos, are summarized in this section.

\subsection{Characteristic quantities}

A number of physical quantities, describing characteristic time and
energy scales of a system under consideration, prove to be especially
important in the theoretical studies of the Loschmidt echo. These
quantities include

\begin{itemize}

\item $g$, {\it mean density of states.} -- By definition, $g(E) =
  \frac{1}{(2\pi\hbar)^{d}} \frac{\rmd V(E)}{\rmd E}$, where $V(E)$ is
  the volume of the phase-space enclosed by a surface of constant
  energy $E$. The mean level spacing $\Delta$ is given by an inverse
  of the density of states, $\Delta = 1/g$. Note that $g$ is the
  smooth part (or average) of the full density of states,
  $g_{\mathrm{full}} (E) = \sum\limits_n \delta(E-E_n)$, where
  $\{E_n\}$ are the eigenenergies of the Hamiltonian.

\item $g_{\mathrm{local}}$, {\it local density of states.} -- Let $|
  \psi \rangle$ be a reference quantum state. The local density of
  states with respect to $| \psi \rangle$ is defined as
  $g_{\mathrm{local}} (E, | \psi \rangle) = \sum\limits_n \big|
  \langle n | \psi \rangle \big|^2 \delta (E-E_n)$, where
  $\{|n\rangle\}$ and $\{E_n\}$ are respectively the eigenstates and
  eigenenergies of the Hamiltonian in question.

\item $N$, {\it effective size of the Hilbert space for a quantum
  state.} -- An initial quantum state $| \psi_0 \rangle$ of a closed
  quantum system can be viewed as a linear combination of $N$
  eigenstates of the Hamiltonian. In a $d$-dimensional system, $N$ is
  given by the volume of the phase space, accessible to $| \psi_0
  \rangle$ in the course of its time evolution, divided by the volume
  of the Planck's cell, $(2\pi\hbar)^d$.

\item $\lambda$, {\it mean Lyapunov exponent.} -- In classical systems
  with chaotic dynamics, a distance between two typical trajectories,
  initiating at infinitesimally close points in phase-space, grows
  exponentially with time. The rate of this growth, averaged over all
  trajectory pairs at a given energy, is the mean Lyapunov exponent
  $\lambda$. The latter characterizes dynamical instability of a
  classical system with respect to perturbations of its initial
  conditions.

\item $t_{\mathrm{E}}$, {\it Ehrenfest time.} -- The center of an
  initially localized wave packet stays close to the phase-space
  trajectory of the corresponding classical particle for times shorter
  than the so-called Ehrenfest time, $t_{\mathrm{E}}$. The latter, in
  chaotic systems, can be estimated as $t_{\mathrm{E}} =
  \frac{1}{\lambda} \ln \frac{L}{\sigma}$, where $\lambda$ is the mean
  Lyapunov exponent of the classical system, $L$ is the characteristic
  linear size of the system, and $\sigma$ is the initial dispersion of
  the wave packet.

\item $t_{\mathrm{H}}$, {\it Heisenberg time.} -- The discreteness of
  the energy spectrum of a closed quantum system becomes dynamically
  important at times comparable to (and longer than) the so-called
  Heisenberg time, $t_{\mathrm{H}} = \hbar/\Delta = \hbar g $, where
  $\Delta$ and $g$ are the mean energy level spacing and the mean
  density of states, respectively. Semiclassical (short-wavelength)
  approximations to quantum dynamics are known to break down beyond
  the Heisenberg time.

\end{itemize}

\subsection{Calculational techniques}

\subsubsection{Semiclassics}
\label{sec:semiclassics}

{\bf Refs.~\cite{JP01Environment, VH03Semiclassical,
    CPJ04Universality, Van04Dephasing, Van06Dephasing, GWG+10Quantum,
    ZOdA11Initial}}

\medskip

\noindent In this context, semiclassics stands for the
short-wavelength approximation to the quantum evolution in terms of
classical trajectories. The key ingredient is the Van Vleck-Gutzwiller
approximation to the propagator (matrix element of the evolution
operator in the position representation)
\begin{equation}
\langle {\bf q}' | e^{-i H t / \hbar} | {\bf q} \rangle = \left(
\frac{1}{2\pi i \hbar} \right)^{d/2} \sum_{\gamma ({\bf q} \rightarrow
  {\bf q}', t)} C_\gamma^{1/2} \exp\left( \frac{i}{\hbar} R_\gamma -
\frac{i \pi}{2} \nu_\gamma \right)
\label{VanVleck-Gutzwiller}
\end{equation}

\begin{itemize}

\item $\gamma$ stands for the classical trajectories going from ${\bf
  q}$ to ${\bf q}'$ in time $t$.

\item $R_\gamma = \int_0^t d\tau \mathcal{L}_{\gamma}$ is the
  Hamilton's principal function for the trajectory $\gamma$. Here,
  $\mathcal{L}_{\gamma}$ denotes the Lagrangian along
  $\gamma$.

\item $C_\gamma = |\det (-\nabla_{{\bf q}'} \nabla_{{\bf q}}
  R_\gamma)|$ is the stability factor of $\gamma$.

\item $\nu_\gamma$ is the number of conjugate points along $\gamma$.

\item $d$ is the number of dimensions of the position space.

\end{itemize}

A combination of Eqs.~(\ref{LE_definition}) and
(\ref{VanVleck-Gutzwiller}) leads to the semiclassical approximation
to the Loschmidt echo,
\begin{equation}
  \includegraphics[width=4in]{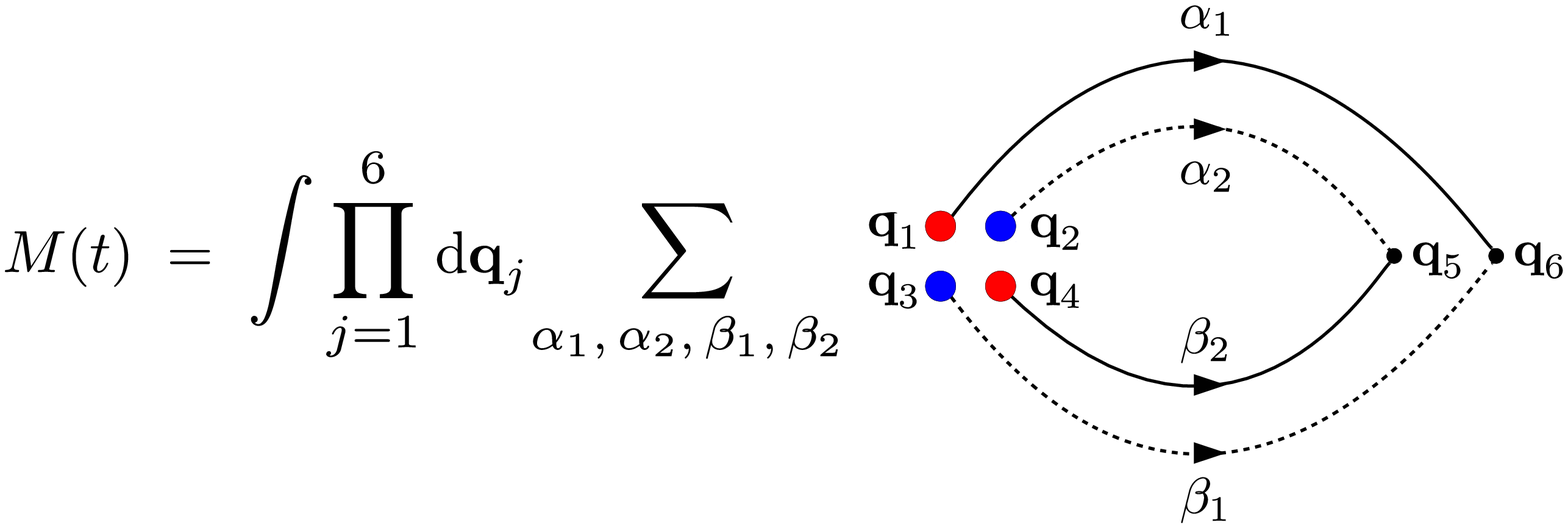}
\label{LE_semiclassics}
\end{equation}
Here, the summand is a product of terms with the following pictorial
representation:

\begin{itemize}

\item A red circles at a point ${\bf q}_j$ corresponds to $\langle
  {\bf q}_j | \psi \rangle$.

\item A blue circles at a point ${\bf q}_j$ correspond to $\langle
  \psi | {\bf q}_j \rangle$. That is, every blue circle is a
  complex conjugate of the corresponding red circle.

\item A solid curve leading from ${\bf q}_j$ to ${\bf q}_k$ along a
  trajectory $\gamma$ corresponds to $(2\pi i \hbar)^{-d/2} \,
  C_{\gamma}^{1/2} \exp \big( i R_{\gamma} / \hbar - i \pi
  \nu_{\gamma} / 2 \big)$. The Hamiltonian $H_1$ is used if $\gamma$
  equals $\alpha_1$, and the Hamiltonian $H_2$ is used if $\gamma$
  equals $\beta_2$.

\item A dashed curve leading from ${\bf q}_j$ to ${\bf q}_k$ along a
  trajectory $\gamma$ corresponds to $(- 2\pi i \hbar)^{-d/2} \,
  C_{\gamma}^{1/2} \exp \big( -i R_{\gamma} / \hbar + i \pi
  \nu_{\gamma} / 2 \big)$. The Hamiltonian $H_1$ is used if $\gamma$
  equals $\alpha_2$, and the Hamiltonian $H_2$ is used if $\gamma$
  equals $\beta_1$. In other words, every dashed curve is a complex
  conjugate of the corresponding solid curve.

\end{itemize}

In Loschmidt echo studies one is typically in interested in the case
of $H_2$ being close to $H_1$. It is common to describe the perturbation
by an operator defined as
\begin{equation}
  \kappa \Sigma = H_2 - H_1 \,,
\label{perturbation}
\end{equation}
where $\kappa$ parametrizes its strength.

The expression in Eq.~(\ref{LE_semiclassics}) is, in general,
extremely complicated to calculate as it involves six spacial
integrals and the sum over four classical trajectories. Semiclassical
evaluations of the Loschmidt echo deal with different approximations
to Eq.~(\ref{LE_semiclassics}).  The following steps are usually
implemented in the semiclassical calculations in order to obtain
meaningful analytical results:

\begin{itemize}

\item[(i)] If the initial state is spatially localized around a point
  ${\bf q}_0$, the diagram of Eq.~(\ref{LE_semiclassics}) can be
  approximated by a simpler diagram in which the points ${\bf q}_1$,
  ${\bf q}_2$, ${\bf q}_3$, and ${\bf q}_4$ merge into ${\bf
    q}_0$. Within this approximation there are only two spacial
  integrals to be done (over ${\bf q}_5$, and ${\bf q}_6$).
 
\item[(ii)] Assuming that the perturbation $\kappa \Sigma$ is
  classically small but quantum mechanically significant (which means
  that the perturbation does not change the topology of the
  trajectories but introduces a phase difference) trajectories
  $\alpha_j$ and $\beta_j$ are identified for $j=1,2$. This is the
  so-called {\bf diagonal approximation}.  The shadowing theorem
  ensures that the identification of classical trajectories $\alpha$
  and $\beta$ is always possible and that the resulting action
  difference can be obtained from the accumulated phase of the
  perturbation along one of the trajectories. For instance, in the
  case where $\Sigma$ only depends on position coordinates
  \begin{equation}
    \label{eq:DeltaRgamma}
    \Delta R_{\gamma} = - \kappa \int_0^t \rmd \tau \ \Sigma({\bf
      q}_{\gamma}(\tau)) \ ,
  \end{equation}
  for $\gamma=\alpha_1,\alpha_2$. The diagonal approximation reduces
  Eq.~(\ref{LE_semiclassics}) to a sum over only two trajectories,
  $\alpha_1$ and $\alpha_2$, of the unperturbed system.

\item[(iii)] A key step that allows to make further progress in the
  semiclassical calculation is to regroup the resulting pair of
  trajectories $(\alpha_1,\alpha_2)$ into two different families:

  \begin{itemize}

  \item Trajectories where ${\bf q}_5$ is near ${\bf q}_6$ and
    $\alpha_1$ evolves close to $\alpha_2$.

  \item The rest of pairs, where the two trajectories are
    uncorrelated.

  \end{itemize}

\end{itemize}

The semiclassical calculation usually proceeds by estimating the
accumulated phases along the trajectories for the two kinds of
resulting pairs, and then performing some kind of average of $M(t)$
(over the perturbation, the initial conditions or the evolution of the
classical trajectories). In the case of complex classical dynamics
these averages are justified and allow to obtain the mean value
$\overline{M(t)}$ of the Loschmidt echo in terms of the main
parameters of problem.

An alternative semiclassical formulation for fidelity amplitude $m(t)$
(defined throught $M(t)=|m(t)|^2$) which avoids the usual 
trajectory-search problem of the standard semiclassics, is the so called 
dephasing representation
\begin{equation}
  \label{eq:odr}
  m(t)=\int \rmd {\bf q} \, \rmd {\bf p} \, \exp\left(-\frac{i}{\hbar}
  \Delta R({\bf q},{\bf p},t)\right) \, W_0({\bf q},{\bf p})
\end{equation}
where 
\begin{equation}
  \label{eq:defWF}
  W_0({\bf q},{\bf p}) = \frac{1}{(2\pi\hbar)^d}\int \rmd
  (\delta\mathbf{q}) \, \exp\left(-\frac{i}{\hbar} \mathbf{p} \cdot
  \delta\mathbf{q}\right) \, \langle {\bf
    q}+\frac{\delta\mathbf{q}}{2}|\psi_0\rangle \langle \psi_0| {\bf
    q}-\frac{\delta\mathbf{q}}{2}\rangle
\end{equation}
is the Wigner function of the initial state
$|\psi_0\rangle$ and
\begin{equation}
  \label{eq:DeltaRdeph}
  \Delta R({\bf q},{\bf p},t)=\int_0^t \rmd \tau [\mathcal{L}_2({\bar
      {\bf q}}(\tau),{\bar {\bf p}}(\tau))-\mathcal{L}_1({\bar {\bf
        q}}(\tau), {\bar {\bf p}}(\tau))]
\end{equation}
is the action difference evaluated along the phase-space trajectory
$\big({\bar {\bf q}}(\tau), {\bar {\bf p}}(\tau) \big)$ evolved from
$\big({\bf q},{\bf p}\big)$ under the unperturbed Hamiltonian
$H_1$. In this way the decay can be attributed to the dephasing
produced by the perturbation of the actions --thus the name dephasing
representation.  In the case where $\Sigma$ only depends on position
coordinates (as in Eq.~(\ref{eq:DeltaRgamma})) the phase difference
(\ref{eq:DeltaRdeph}) reads
\begin{equation}
  \Delta R({\bf q},{\bf p},t) = - \kappa \int_0^t \rmd \tau
  \ \Sigma({\bar {\bf q}}(\tau)) \ .
\end{equation}
For generic chaotic systems and initially localized states, the
calculation of $M(t)$ using the dephasing representation follows the
same lines of the standard semiclassics. The standard semiclassical
approach becomes an initial-value problem once the integration over
the final position is traded by the integration over the initial
momentum using the stability factor $C_{\gamma}$ as the Jacobian of
the transformation. However, Eq.~(\ref{eq:odr}) bears the advantage
that it regards $m(t)$ as the solution of an initial-value problem, as
opposed to a boundary-value problem. This proves especially convenient
in situations when the sums over classical trajectories are evaluated
explicitly.

\subsubsection{Random matrix theory}

{\bf Refs.~\cite{CLM+02Measuring, CT03uniform, GPS04random,
    SS04Recovery, SS05Fidelity, KSP+08Surprising, KNS11Parametric,
    KR12Fidelity}}

\medskip

\noindent Random matrix theory is a powerful technique to understand
the statistical behavior of quantum complex systems. Among the latter
quantum systems exhibiting chaotic dynamics in the classical limit are
particularly important, as their spectral properties are well
described by averages taken over appropriate ensembles of matrices
that satisfy certain general symmetry restrictions. The simplest
Hamiltonian ensemble is that of real symmetric $N \times N$ matrices
(appropriate for cases with time-reversal symmetries) usually referred
to as Gaussian Orthogonal Ensemble.

The invariance of the probability distribution under orthogonal 
transformations and the statistical independence of the matrix elements 
lead to a joint probability distribution of the matrix elements of the 
Hamiltonian,
\begin{equation}
 P_N(H)=K_N \exp\left(-\frac{\tr{\{H^2\}}}{4 v}\right), 
\end{equation}
where $v$ defines the scale of the matrix elements and $K_N$ is a 
normalization constant.
Each of the $N(N+1)/2$ independent matrix element $H_{ij}$ is a zero-centered 
Gaussian such that
\begin{eqnarray}
\overline{H_{ij}}=& 0 & i \leq j, \nonumber \\
\overline{H^2_{ij}}=& (1+\delta_{i j}) v^2 & i \leq j \nonumber .
\end{eqnarray}

The averages are taken over the matrix ensemble. The Loschmidt echo in
complex systems can be addressed by imposing the Random Matrix
hypothesis for the forward evolution Hamiltonian $H_1$, for the
perturbation $\kappa \Sigma$ representing the imperfection of the
time-reversal, or for both. Generically, the same results are
obtained, independently of the choice of the matrix considered to be
random. Averages of the Loschmidt echo over matrix ensembles are
justified for ergodic classical dynamics. The absence of finite
classical time scales, like the inverse Lyapunov exponent or the
escape rate, in the Random matrix theory hinders the description of
the Loschmidt echo decays that depend on those quantities.

Recently, using the Random-Matrix-Theory approach, Kohler and
coworkers have established relations between the averaged fidelity
decay and the so-called cross-form factor, characterizing parametric
correlations of the energy spectra of the unperturbed and perturbed
systems. Notably, these relations exist not only in the case of fully
chaotic unperturbed and perturbed systems, but also in the case of a
regular system perturbed by a chaotic perturbation.

\subsubsection{Numerical simulations}

{\bf Refs.~\cite{TK84accurate, Rae96Computer, DBS11Efficiency}}

\medskip

\noindent A numerical simulation of the time-evolution of initially
localized wave-packets under slightly different Hamiltonians, $H_1$
and $H_2$, is a valuable tool in understanding the behavior of the
Loschmidt echo in different systems and various regimes. Even though
there exist numerous approaches to solving the time-dependent
Schr{\"o}dinger equation on a computer, the majority of numerical
studies of the Loschmidt echo have been concerned with the following
two methods. Both methods provide accurate, efficient, and stable
approximations to the evolution operator $e^{-i A \tau}$, where the
operator $A$ corresponds a (properly rescaled) Hamiltonian, and $\tau$
denotes a sufficiently short propagation time-step.

\begin{itemize}

  \item {\it Trotter-Suzuki algorithm.} -- This method involves three
    implementation stages. First, one decomposes $A$ into a finite
    (and practically small) number of components, $A = A_1 + A_2 +
    \ldots + A_n$, such that the operator $e^{-i A_j t}$ can be
    constructed analytically for all $j = 1,2,\ldots,n$. Second, one
    defines the symmetric operator $U_2(\tau) = e^{-i A_n \tau/2}
    \ldots e^{-i A_2 \tau/2} e^{-i A_1 \tau} e^{-i A_2 \tau/2} \ldots
    e^{-i A_n \tau/2}$. Third, one constructs the operator $U_4(\tau)
    = U_2(p\tau) U_2(p\tau) U_2((1-4p)\tau) U_2(p\tau) U_2(p\tau)$
    with $p = 1/(4-4^{1/3})$. This provides a unitary approximation to
    the original propagator accurate up to the 4th order in $\tau$,
    i.e., $e^{-i A \tau} = U_4(\tau) + \mathcal{O}(\tau^5)$.

  \item {\it Chebyshev-polynomial expansion.} -- This approach
    requires that the operator $A$ is normalized in such a way that
    all its eigenvalues lie in the interval between -1 and 1. The
    method involves constructing the operator $S_N (\tau) = J_0(\tau)
    + 2 \sum_{n=1}^N (-i)^n J_n(\tau) T_n(A)$, where $\{ J_n \}$ and
    $\{ T_n \}$ are, respectively, the Bessel functions of the first
    kind and the Chebyshev polynomials of the first kind. In actual
    computations, one makes use of the recurrence relation $T_{n+1}(A)
    = 2 A T_n(A) - T_{n-1}(A)$, with $T_0(A) = 1$ and $T_1(A) = A$, to
    calculate $S_N(\tau)$. Then, for large values of $n$ and fixed
    $\tau$, the Bessel function $J_n(\tau)$ rapidly decays as a
    function of $n$. In fact, $|J_n(\tau)| \sim (\tau/2)^n / n!$ as $n
    \rightarrow \infty$. This is why $S_N(\tau)$, with a sufficiently
    large $N$, provides an extremely accurate approximation to the
    propagator in question, i.e., $e^{-i A \tau} = S_N(\tau) +
    \mathcal{O}\big( (\tau/2)^N / N!  \big)$.

\end{itemize}

\subsection{Decay of the Loschmidt echo: different regimes}
\label{results}

The decay of the Loschmidt echo mainly depends upon the underlying
classical dynamics of the system, the initial state, and the nature
and strength of the perturbation $\kappa \Sigma$.

The behavior of the Loschmidt echo is fairly well understood for
single-particle quantum systems whose dynamics is fully chaotic in the
classical limit. Some progress has been done towards the theory of the
Loschmidt echo in systems with regular and mixed phase space.

\subsubsection{Chaotic dynamics}
\label{regimes_ch}

{\bf Refs.~\cite{JP01Environment, JSB01Golden, CT02Sensitivity,
    WC02Quantum, CT03uniform, Wis03Short, CPJ04Universality,
    GPS04random, GG09Long}}

\medskip

\begin{figure}[ht]
\center
\includegraphics[width=3.7in]{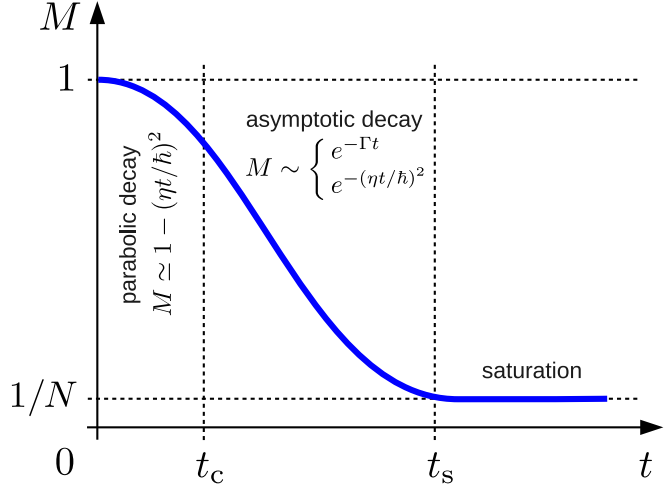}
\caption{Decay regimes of the Loschmidt echo in quantum systems with
  chaotic classical limit. Characteristic crossover times are
  $t_\mathrm{c} \simeq \hbar / \eta$, signaling the end of the initial
  parabolic decay, and $t_\mathrm{s}$ ($\simeq \Gamma^{-1} \ln N$ for
  the non-perturbative/exponential regime) indicating the onset of the
  saturation.}
\label{fig.time}
\end{figure}

\noindent The time-decay of the Loschmidt echo, averaged over an
ensemble of initial states, Hamilton operators, or perturbations,
typically exhibits three consecutive stages (see Fig. \ref{fig.time}):
\begin{itemize}

\item {\bf Short-time parabolic decay}, $\overline{M(t)} \simeq 1 -
  (\eta t/\hbar)^2$. It is a short, initial stage of the Loschmidt echo
  decay in all quantum systems. Here, $\eta$ is an average
  dispersion of the perturbation operator evaluated with respect to
  the initial state, i.e., $(\eta / \kappa)^2 = \overline{\langle
    \psi | \Sigma^2 | \psi \rangle} - \overline{\langle \psi | \Sigma
    | \psi \rangle^2}$. The parabolic decay holds for times $t$ short
  enough for the propagators of the unperturbed and perturbed systems,
  $\exp\left( -i H_j t /\hbar \right)$ with $j=1,2$, to be reliably
  approximated by their second order Taylor expansions, $1 - i H_j t /
  \hbar - (H_j t)^2 / (2 \hbar^2)$.

\item {\bf Intermediate-time asymptotic decay}. The short-time
  parabolic decay is typically followed by an ``asymptotic'' decay
  regime, whose functional form depends on the strength of the
  perturbation.

  \begin{itemize}

    \item {\bf Perturbative/Gaussian regime}, $\overline{M(t)} \simeq
      \exp(-(\eta t/\hbar)^2)$. This decay holds for ``weak''
      Hamiltonian perturbations, such that the absolute value of a
      characteristic matrix element of the perturbation operator is
      small compared to the mean energy level-spacing of the
      unperturbed Hamiltonian. Since $\eta \sim \kappa$, the decay
      rate is quadratic in the perturbation strength.

    \item {\bf Non-perturbative/Exponential regime}, $\overline{M(t)}
      \simeq \exp(-\Gamma t)$. This non-perturbative decay regime is
      typically observed for stronger Hamiltonian perturbations, i.e.,
      for perturbations large on the scale of the mean level spacing
      of the unperturbed Hamiltonian. The functional form of the
      dependence of the decay rate $\Gamma$ on the perturbation
      strength $\kappa$ is different for ``global'' and ``local''
      Hamiltonian perturbations (see below).

  \end{itemize}

\item {\bf Long-time saturation}, $\overline{M(t)} \sim N^{-1}$. At
  long times, the asymptotic decay of the Loschmidt echo is followed
  by a saturation (or freeze) at a value inversely proportional to the
  size $N$ of the effective Hilbert space of the system. $N$ is given
  by the volume of the phase space, that is available to the state of
  the system in the course of its time evolution, divided by the
  volume of the Planck's cell, $(2\pi\hbar)^d$. The value, at which the 
  Loschmidt echo saturates, is independent of the perturbation
  strength.

  An explicit expression is available for the saturation value of the
  Loschmidt echo in two-dimensional quantum billiards, with strongly
  chaotic dynamics in the classical limit. Thus, if the initial state
  of the particle is given by the Gaussian wave function $\psi({\bf
    q}) = (\pi \sigma^2)^{-1/2} \exp\big(i {\bf p}_0 ({\bf q}-{\bf
    q}_0) / \hbar - ({\bf q}-{\bf q}_0)^2/(2 \sigma^2) \big)$ with
  ${\bf q}_0$, ${\bf p}_0$, and $\sigma$ denoting respectively the
  average position, momentum, and position uncertainty of the
  particle, and the area of the billiard is $A$, then the long-time
  saturation of the average Loschmidt echo is given by
  $\overline{M(t)} \simeq \sqrt{2\pi} \, \hbar \sigma / (|{\bf p}_0|
  A)$. This formula holds in the semiclassical regime, such that
  $\hbar/|{\bf p}_0| \ll \sigma \ll \sqrt{A}$.
  
\end{itemize}

\begin{figure}[ht]
\center
\includegraphics[width=3in]{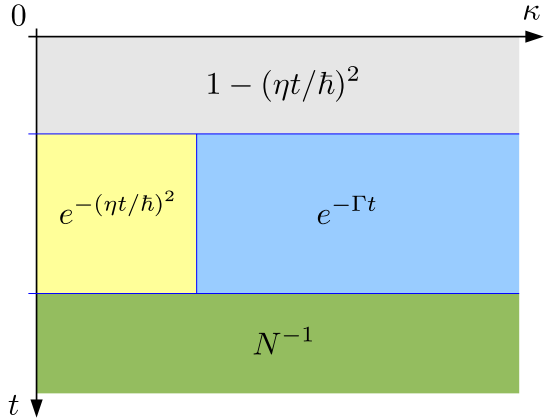}
\caption{Typical regimes of the Loschmidt echo decay in chaotic
  systems.}
\label{fig-decay_regime_table}
\end{figure}

The above classification is schematically illustrated in
Fig.~\ref{fig-decay_regime_table}.

\bigskip

\noindent {\it Global perturbations}

\smallskip

\noindent {\bf Refs.~\cite{JP01Environment, JSB01Golden,
    CPW02Decoherence, CPJ04Universality, PZ02Stability}}

\medskip

\noindent A Hamiltonian perturbation is said to be ``global'' if it
affects all (or a dominant part) of the phase-space that is accessible
to the system in the course of its time-evolution. The accessible
phase-space depends on the initial state of the system.

\begin{figure}[ht]
\center
\includegraphics[width=3in]{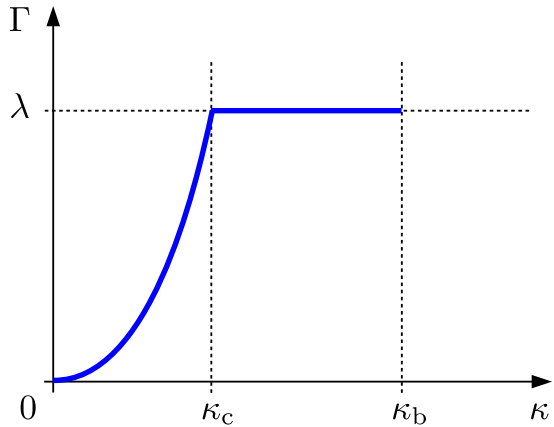}
\caption{Exponential decay rate $\Gamma$ as a function of the
  perturbation strength $\kappa$ in the case of a global Hamiltonian
  perturbation.}
\label{fig-Gamma_vs_kappa-global}
\end{figure}

For global perturbations, the dependence of the decay rate $\Gamma$ on
the perturbation strength $\kappa$ can be approximated by a
piecewise-continuous function, shown schematically in
Fig.~\ref{fig-Gamma_vs_kappa-global}. More precisely, the
semiclassical analysis of the exponential decay regime of the
Loschmidt yields
\begin{equation}
\label{eq:dec_LE}
\overline{M(t)}= e^{-c \kappa^2 t} +  b(\kappa,t) \ e^{-\lambda t} \ .
\end{equation}
The first term in Eq.~(\ref{eq:dec_LE}) stems from uncorrelated pairs
$(\alpha_1,\alpha_2)$ of contributing trajectories, while the for
second term is a contribution of correlated trajectory pairs, such
that $\alpha_1$ evolves close to $\alpha_2$ (see
Sec.~\ref{sec:semiclassics}). Unlike the constant $c$, the prefactor
$b$ generally exhibits an explicit dependence on both $t$ and
$\kappa$. However, this dependence is sub-exponential, and, as a
consequence, the decay rate $\Gamma$ of the Loschmidt echo is
dominated by the minimum of $c \kappa^2$ and $\lambda$, leading to a
commonly accepted (but oversimplified) interpretation of the
exponential decay (see Fig.~\ref{fig-Gamma_vs_kappa-global}).

For perturbation strengths $\kappa$ that are weak compared to a
critical strength $\kappa_{\mathrm{c}}$, the dependence is parabolic,
$\Gamma = c \kappa^2$.  The semiclassical calculations give the
expression of $c$ in terms of correlation functions of the
perturbation, and the random matrix approach leads to the general
dependence $c \kappa^2=2 \pi \eta^2/(\hbar \Delta)$, where $\eta$ is
given by the mean value of the width of the local density of states
with respect to the initial state $|\psi_0\rangle$. Such a behavior is
commonly referred to as the {\bf Fermi-golden-rule regime}.

For perturbations that are stronger than $\kappa_{\mathrm{c}}$, but
weaker than a certain bound value $\kappa_{\mathrm{b}}$, the decay
rate $\Gamma$ is independent of $\kappa$ and equals the average
Lyapunov exponent $\lambda$ of the underlying classical system. This
decay is known as the {\bf perturbation-independent} or {\bf Lyapunov
  regime}. This regime holds for perturbations up to the breakdown
strength $\kappa_{\mathrm{b}}$, beyond which it is no longer possible
to regard every trajectory of the perturbed system as a result of a
continuous deformation of the corresponding unperturbed trajectory, so
that bifurcations have to be taken into account. The Lyapunov regime
is especially remarkable, as the decay rate is totally independent of
the strength of the perturbation which is at the origin of the decay.

The exponential decay of the Loschmidt echo is generally followed in
time by a saturation at a value on the order of $N^{-1}$, where $N$ is
the size of the effective Hilbert space. This implies that, in the
non-perturbative regime, the saturation occurs at time $t_{\mathrm{s}}
\simeq \Gamma^{-1} \ln N$. In the case of a two-dimensional chaotic
billiard (introduced above), the saturation time is given by
$t_{\mathrm{s}} \simeq \Gamma^{-1} \ln\big( |{\bf p}_0| A / (\hbar
\sigma) \big) - (2\Gamma)^{-1} \ln(2\pi)$. It is interesting to note
that the saturation time can, in principle, be arbitrarily long. In
particular, $t_{\mathrm{s}}$ can exceed other important time scales of
quantum dynamics, such as the Heisenberg time $t_{\mathrm{H}} \equiv
\hbar g(E)$ with $g(E)$ denoting the density of states. Indeed, in
two-dimensional billiards the saturation time $t_{\mathrm{s}}
\rightarrow \infty$ as $|{\bf p}_0|\rightarrow \infty$, while the
Heisenberg time $t_{\mathrm{H}}$ is independent of the particle's
momentum.

The exponential decay, $\overline{M(t)} \simeq \exp(-\Gamma t)$, only
holds for perturbations weaker than $\kappa_{\mathrm{b}}$. Beyond this
threshold, the exponential regime breaks down and gives way to another
regime, in which the Loschmidt echo exhibits a Gaussian dependence on
time.

Some of the mentioned regimes can be obtained using alternative
approaches relating the Loschmidt echo with the two-point time
auto-correlation function of the generator of the perturbation.

\bigskip

\noindent {\it Local perturbations}

\smallskip

\noindent {\bf Refs.~\cite{GR07Loschmidt, GWRJ08Loschmidt,
    HKS08Algebraic, AW09Loschmidt, KKS+11Fidelity}}

\medskip

\noindent A Hamiltonian perturbation is said to be ``local'' if it is
concentrated in a small region of the phase space accessible to the
system. In the case of chaotic dynamics, the phase space extent of a
local perturbation can be characterized by a rate $\gamma$, known as
the ``escape'' rate, that is defined as the rate at which trajectories
of the corresponding classical system visit the perturbation
region. For local perturbations, the escape rate is small compared to
the characteristic rate at which a typical trajectory ``explores'' the
dynamically available phase space, and also small compared to the
average Lyapunov exponent of the system.

\begin{figure}[ht]
\center
\includegraphics[width=3in]{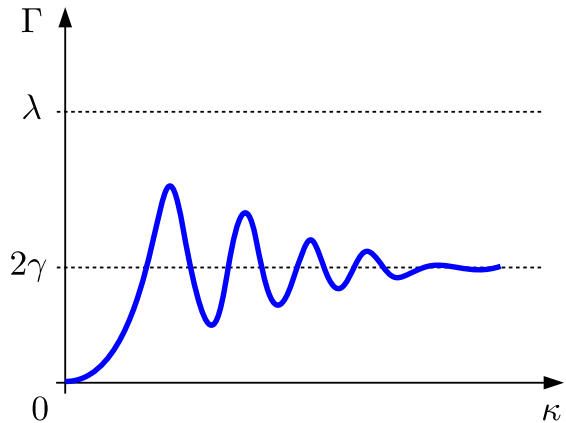}
\caption{Exponential decay rate $\Gamma$ as a function of the
  perturbation strength $\kappa$ in the case of a local Hamiltonian
  perturbation.}
\label{fig-Gamma_vs_kappa-local}
\end{figure}

Figure~\ref{fig-Gamma_vs_kappa-local} illustrates the characteristic
dependence of the decay rate $\Gamma$ on the perturbation strength
$\kappa$. The function $\Gamma(\kappa)$ exhibits a number of
distinctive features. Thus, for sufficiently weak perturbations, the
decay rate grows quadratically with the perturbations strength,
$\Gamma \sim \kappa^2$, demonstrating the {\bf Fermi-golden-rule
  regime}. In the limit of strong perturbations, $\kappa \rightarrow
\infty$, the decay rate $\Gamma$ saturates at a
perturbation-independent value $2\gamma$. This saturation is known as
the {\bf escape-rate regime}. The crossover from the Fermi-golden-rule
to the escape rate regime is non-monotonic, and 
$\Gamma(\kappa)$ generally exhibits well-pronounced {\bf
  oscillations}. The amplitude and frequency of these oscillations
depend on the nature and physical properties of the particular
Hamiltonian perturbation.  An important point is that the function 
$\Gamma(\kappa)$ is given for
all strengths $\kappa$ by the width of the local density of states.

The exponential decay, $\overline{M(t)} \simeq \exp(-\Gamma t)$ with
the non-monotonic function $\Gamma(\kappa)$ described above and
illustrated in Fig.~\ref{fig-Gamma_vs_kappa-local}, has been obtained
by semiclassical analysis. (Its existence has now been observed in
numerical and laboratory experiments). The key assumption of any
semiclassical reasoning is that the de Broglie wavelength corresponds
to the shortest length scale of the system. This assumption is
obviously violated for systems, in which the spatial linear size of
the Hamiltonian perturbation is small compared to the de Broglie
wavelength. Generally, the decay of the Loschmidt echo in such systems
is different from the exponential decay. For instance, two-dimensional
chaotic systems in the limit of point-like Hamiltonian perturbations,
for which $\gamma \rightarrow 0$, exhibit the algebraic,
inverse-quadratic decay of the Loschmidt echo, $\overline{M(t)} \sim
t^{-2}$.

\subsubsection{Regular and mixed dynamics}
\label{regimes_regmix}

\smallskip

\noindent {\bf Refs.~\cite{Pro02General, PZ03Quantum, JAB03Anomalous,
    SL03Recurrence, WH05Quantum, WCL07Stability, Gou11Nonmonotonic}}

\medskip

\noindent Time dependence of the Loschmidt echo in quantum system with
regular or mixed classical dynamics is typically more complex than
that in fully chaotic systems. As a result of this complexity, a
comprehensive classification of decay regimes is still
lacking. Nevertheless, a number of important results have been
established for the Loschmidt echo decay in systems whose phase space
is predominantly regular in the classical limit.

In the limit of weak Hamiltonian perturbations, $\kappa \rightarrow
0$, the average Loschmidt echo generally exhibits a {\bf Gaussian
  decay}. The duration $t_{\mathrm{G}}$ of this decay is inversely
proportional to the perturbation strength, $t_{\mathrm{G}} \sim
\kappa^{-1}$. The existence of the Gaussian decay requires the
perturbation to be sufficiently weak such that the decay time
$t_{\mathrm{G}}$ is long in comparison with any relaxation (averaging)
time scale of the system.

If the Hamiltonian perturbation is sufficiently strong and varies
rapidly (or is almost ``random'') along a typical classical trajectory
of the unperturbed Hamiltonian, then the average Loschmidt echo is
known to exhibit the {\bf algebraic decay} $\overline{M(t)} \sim
t^{-3d/2}$, with $d$ being the dimensionality of the
system. This power-law decay is faster than the decay of the overlap
of the corresponding classical phase-space densities, $t^{-d}$.

Numerous case studies of the unaveraged (individual-realization)
Loschmidt echo $M(t)$ have revealed that the functional form of the
decay depends strongly upon the location of the initial quantum state
with respect to the phase space of the underlying, unperturbed and
perturbed, classical systems. In addition, the decay is sensitive to
certain properties of the Hamiltonian perturbation. For instance, the
Loschmidt echo decays differently depending on whether the
perturbation is global or local in phase space, whether it vanishes
under averaging over time, or whether it preserves or destroys
integrability of the underlying classical system. A number of
different regimes, such as the {\bf exponential decay} and power-law
decays with varies decay exponents, have been observed in numerical
simulations. Interesting phenomena of {\bf quantum revivals}, closely
related to (quasi-)periodicity of the underlying classical motion, and
temporary {\bf quantum freeze} of the Loschmidt echo have been
reported in various studies. However, a comprehensive classification
of all decay regimes of the Loschmidt echo in regular systems still
remains a challenge in the field.

\subsubsection{Numerical observation of various decay regimes}

{\bf Refs.~\cite{JSB01Golden, CLM+02Measuring, CPW02Decoherence,
    CPJ04Universality, GR07Loschmidt, GWRJ08Loschmidt, AW09Loschmidt,
    GW11Loschmidt}}

\medskip

\begin{figure}[ht]
\center
\includegraphics[width=4in]{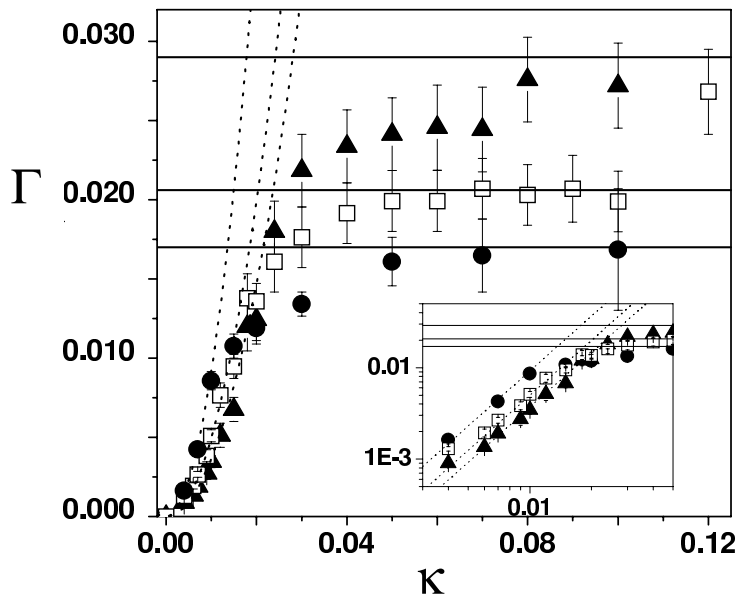}
\caption{Decay rate of the Loschmidt echo as a function of the
  perturbation strength $\kappa$ for the three concentrations of
  impurities in a Lorentz gas. The rates $\Gamma$ are normalized to
  the group velocity of the initial wave packet. The solid lines are
  the corresponding classical Lyapunov exponents and the dashed lines
  are fits to the quadratic behavior. Inset: a log-log scale of the
  same data showing the quadratic increase of $\Gamma$ for small
  perturbations. Adapted from Ref.~\cite{CPJ04Universality}. Copyright
  (2004), American Physical Society.}
\label{fig-Gamma_vs_kappa-lorenz}
\end{figure}

\begin{figure}[ht]
\center
\includegraphics[width=3.5in]{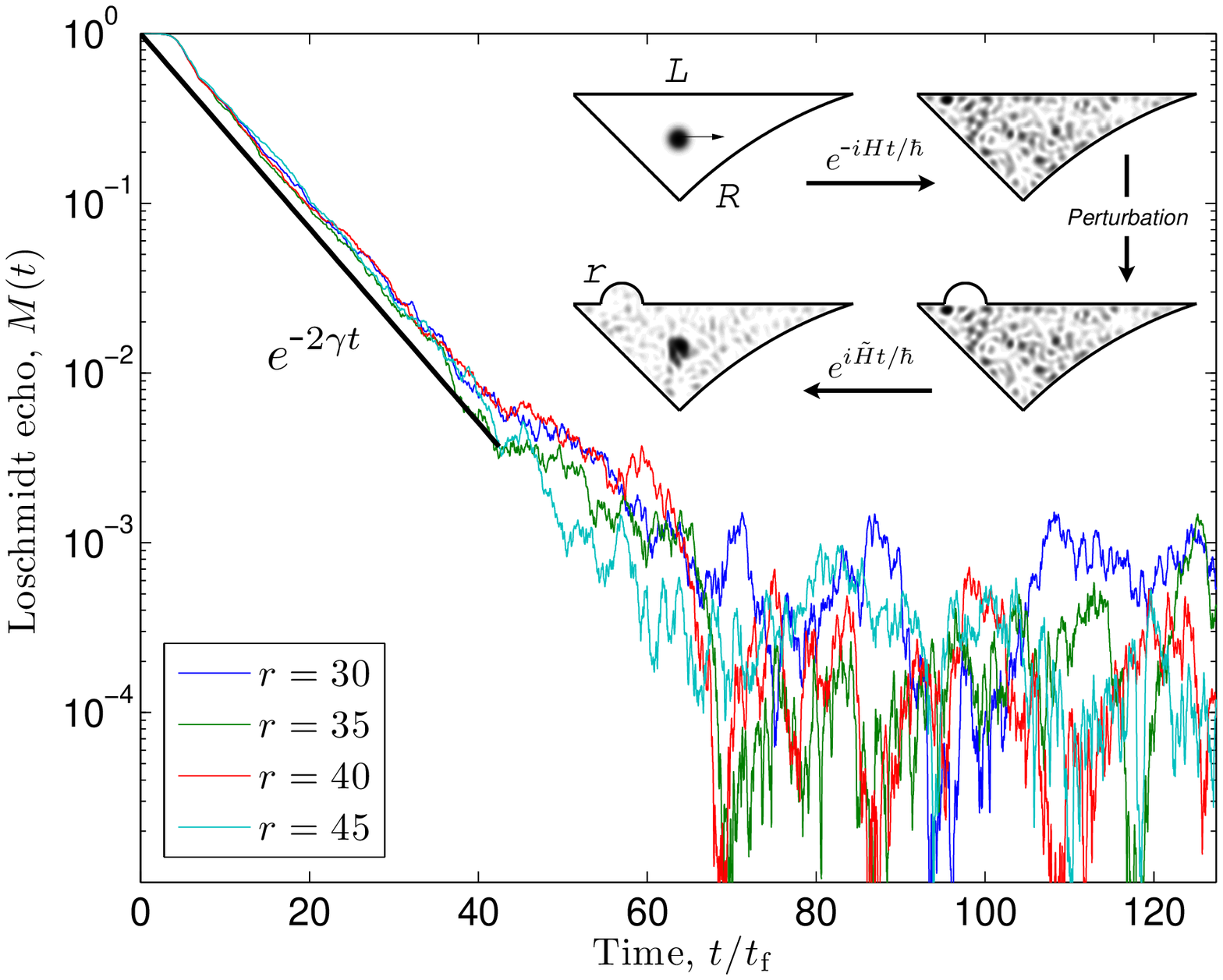}
\caption{The Loschmidt echo decay in the desymmetrized diamond
  billiard for four different values of the curvature radius $r$ of
  the arc boundary deformation. The solid straight line gives the
  trend of the $\exp(-2\gamma t)$ decay. Inset: Forward-time wave
  packet evolution in the unperturbed billiard, followed by the
  reversed-time evolution in the perturbed billiard. The arrow shows
  the momentum direction of the initial wave packet. The propagation
  time corresponds to approximately 10 collisions of the classical
  particle. Adapted from Ref.~\cite{GR07Loschmidt}. Copyright (2007),
  American Physical Society.}
\label{fig-blocal}
\end{figure}

\noindent Numerical simulations have been of foremost importance for
establishing the results summarized in sections \ref{regimes_ch} and
\ref{regimes_regmix}.  The main body of numerical work has been
devoted to one-particle systems. Various chaotic dynamical systems,
globally perturbed, have been studied numerically; among them the
Lorentz gas, two-dimensional hard-wall and soft-wall billiards. As a
prominent example, Fig. \ref{fig-Gamma_vs_kappa-lorenz} shows the
crossover from the perturbative Fermi-golden-rule regime to the
Lyapunov regime of the decay rate of the Loschmidt echo as a function
of the perturbation strength in the Lorentz gas. The predictions for
local perturbations, described in Sec.~\ref{regimes_ch}, were also
tested in a chaotic billiard system. In this case, a deformation of a
small region of the boundary was considered as the perturbation (see
Fig.~\ref{fig-blocal}).

\begin{figure}[ht]
\center
\includegraphics[width=4in]{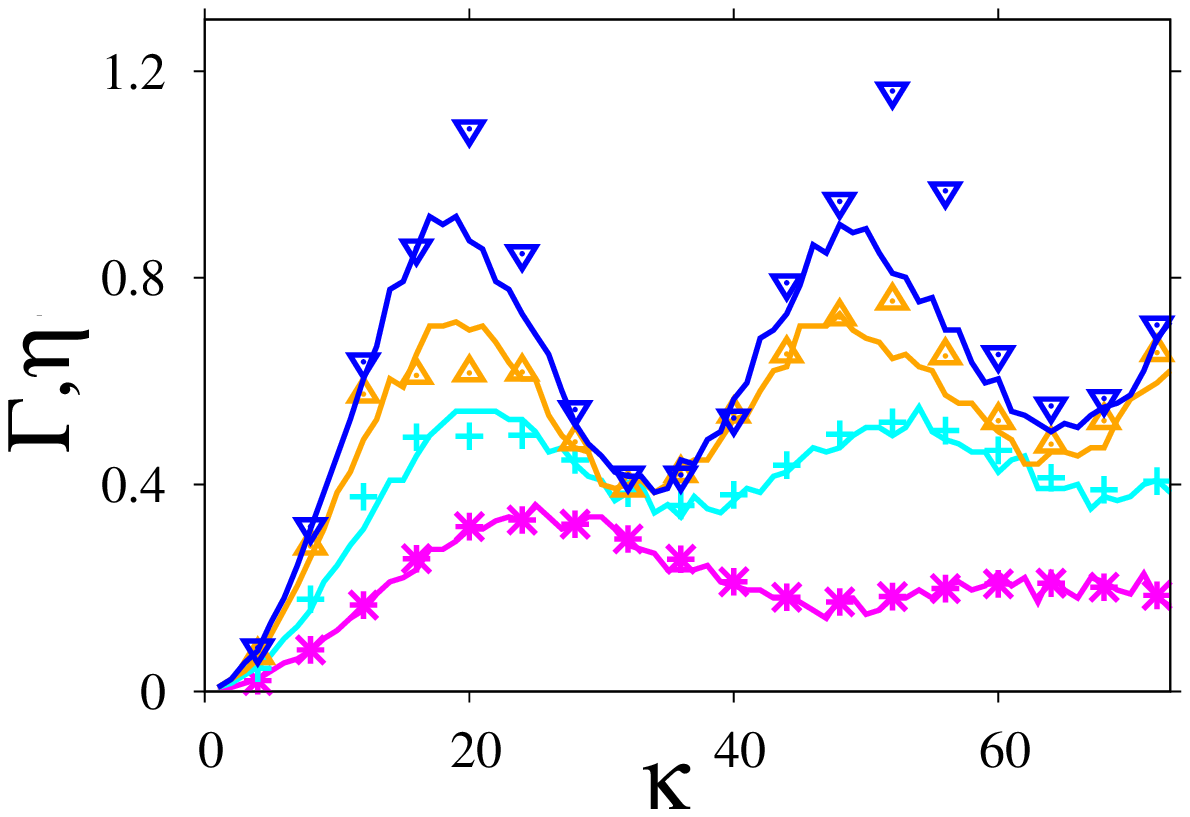}
\caption{Decay $\Gamma$ of the Loschmidt echo of a cat map perturbed
  locally with a shear in momentum as a function of the scaled
  perturbation strength $\kappa$. The symbols correspond to the
  following fraction of the phase space that is perturbed:
  $\bigtriangledown$ ($40 \%$), $\bigtriangleup$ ($30 \%$), $+$ ($20
  \%$) and $\ast$ ($10 \%$). The width $\eta$ of the local density of
  states is also plotted with solid lines. This width gives the spread
  of the states of the unperturbed system in the bases of the
  perturbed one It is noticeable that, the smaller is the perturbed
  region, the more similar to $\eta$ $\Gamma$ is. Adapted from
  Ref.~\cite{AW09Loschmidt}. Copyright (2009), American Physical
  Society.}
\label{fig-Gamma_vs_kappa-cat}
\end{figure}

Other enlightening numerical studies were carried in quantum
maps. These systems possess all the essential ingredients of the
chaotic dynamics and are, at the same time, extremely simple from a
numerical point of view, both at the classical and at the quantum
levels. The predictions for locally perturbed chaotic systems were
also observed in the paradigmatic cat map.
Figure~\ref{fig-Gamma_vs_kappa-cat} shows the decay rate of the
Loschmidt echo as a function of the perturbation strength $\kappa$ in
the cat map under the action of a local perturbation. When the map was
perturbed in all its phase space, three decay regimes (instead of two)
were observed depending on the perturbation strength: (i) the
Fermi-golden-rule regime for weak perturbations, (ii) a regime of an
oscillatory dependence of the decay rate on $\kappa$ at intermediate
perturbation strengths, and (iii) the Lyapunov regime for strong
perturbations. The complete understanding of the crossover between the
last two regimes remains an open problem.

\subsubsection{Beyond the ``standard'' picture:  fluctuations, disorder, many-body systems}

{\bf Refs.~\cite{AGM03Loschmidt, STB03Hypersensitivity,
    PJ05Mesoscopic, MH06Loschmidt, QSL+06Decay, PD07Fidelity,
    MH08Fidelity, ZDL+Quantifying}}

\medskip

\noindent Decay regimes of the average Loschmidt echo, discussed
above, are the ones most commonly observed, and generally regarded as
``standard'', in low-dimensional quantum systems. However, when
analyzing the Loschmidt echo beyond a simple averaging over the
initial state or perturbation, or in a more complex setup, one often
encounters departures from the standard picture.

It was fist pointed out by Silvestrov and coworkers that, for
sufficiently strong perturbations and on time scales short compared to
the Ehrenfest time, statistical fluctuations generally play an
important role in the problem of the Loschmidt echo decay. In
particular, the average Loschmidt echo, $\overline{M(t)}$, is
typically dominated by rare fluctuations, characteristic of only a
small fraction of the chaotic phase space. In the main part of the
phase space, however, the unaveraged Loschmidt echo, $M(t)$, decays
much faster than $\overline{M(t)}$: this decay can be as fast as
double-exponential, $M(t) \sim \exp(-\mathrm{constant} \times
e^{2\lambda t})$. It is only after the Ehrenfest time that $M(t)$
follows $\overline{M(t)}$ in the main part of the phase space.

The variance of the Loschmidt echo, $\overline{M^2} - \overline{M}^2$,
was addressed by Petitjean and Jacquod within both a semiclassical and
a Random Matrix Theory approach. The variance was shown to exhibit a
rich nonmonotonous dependence on time, characterized by an algebraic
growth at short times followed by an exponential decay at long times.

In disordered systems, the Loschmidt echo may exhibit even richer
decay than that
%Decay regimes of the Loschmidt echo in disordered systems are even
%richer than those
in low-dimensional dynamical systems, since the elastic mean-free path
and the disorder correlation length enter as relevant scales. For
long-range disorder (small-angle scattering) an asymptotic regime
governed by the classical Lyapunov exponent emerges. A Josephson flux
qubit operated at high energies leads to a chaotic dynamics, and the
Lyapunov regime can be obtained when a Loschmidt setup is considered
in such a system.

The Loschmidt echo studies in many-body systems are comparatively less
developped than those in one-body cases. The numerical calculation of
the many-body Loschmidt echo is a highly demanding computational task,
and the standard approximations are in difficulty for describing the
small difference obtained after the forward and backward
time-evolutions. For this reason, the most reliable results of
Loschmidt echo in many-body systems are those of a one dimensional
spin chain or in an Ising model with transverse field, where the
influence of criticality in the fidelity decay has been
established. Trapped Bose-Einstein condensates have been treatead as a
many-body Loschmidt echo setup using mean-field approaches.

\subsection{Phase-space representation and classical fidelity}

{\bf Refs.~\cite{BC02Quantum, BCV03Stability, Eck03Echoes,
    GSS03Classical, CPJ04Universality, VP04Faster}}

\medskip

\noindent An intuitively appealing representation of the Loschmidt
echo in phase space is obtained by expressing
Eq.~(\ref{LE_definition}) as
\begin{equation}
M(t) =
\frac{1}{(2\pi\hbar)^d}\int\mathrm{d}\mathbf{q}\int\mathrm{d}\mathbf{p}
\ W_{H_1}(\mathbf{q},\mathbf{p};t)\ W_{H_2}(\mathbf{q},\mathbf{p};t)
\ , \label{eq-LEWigner}
\end{equation}
where $W_{H_1}$ and $W_{H_2}$ are the Wigner functions resulting from
the evolution of $W_0(\mathbf{q},\mathbf{p})$, introduced in
Eq.~(\ref{eq:defWF}), under the action of the Hamiltonian operators
$H_1$ and $H_2$ respectively. This approach is naturally connected to
the dephasing representation of Sec.~\ref{sec:semiclassics} and
provides a particularly useful framework in the study of decoherence,
as the Wigner function is a privileged tool to understand the
connection between quantum and classical dynamics.

An initial Gaussian wave-packet is associated with a Gaussian Wigner
function that will develop in phase-space non-positive structures and
be deformed under the evolution of $H_1$ and $H_2$. The sensitivity of
the overlap (\ref{eq-LEWigner}) to the non-positive parts and to the
deformations of the Wigner function can be related with the different
decay regimes discussed in Sec.~\ref{regimes_ch}.

The form (\ref{eq-LEWigner}) of the quantum fidelity allows to define
a classical fidelity by using in a classical problem the Liouville
distributions $L_{H_1}$ and $L_{H_2}$ instead of $W_{H_1}$ and
$W_{H_2}$ respectively. This definition follows from the logical
representation of the Liouville distribution as the classical limit of
the Wigner function. However, the definition by an overlap of
distributions does not convey the sense of classical reversibility: a
trajectory evolving forward in time with $H_1$ and backwards with
$H_2$ would give a considerable contribution to the overlap not only
if it ends up close to the initial point, but whenever the final point
is in the neighborhood of those of the initial distribution. The
labeling of trajectories and particles, characteristic of classical
mechanics, is then not taken into account in this definition of
classical fidelity.

The correspondence principle dictates that the quantum fidelity
follows its classical counterpart up to the Ehrenfest time. However,
such a correspondence does not imply that, in the intermediate-time
asymptotic decay of globally perturbed chaotic systems, the Lyapunov
regime is only present up to this characteristic time scale. In fact,
the Lyapunov regime is not simply an effect of the classical-quantum
correspondence. For instance, in a two-dimensional billiard, the
saturation time $t_{\rm S}$ (see Sec.~\ref{regimes_ch}) that would
signal the end of the asymptotic decay (and so the end of the Lyapunov
decay if one is in the appropriate regime) can be much larger than the
Ehrenfest time.

\section{Experiments}

\subsection{Nuclear Magnetic Resonance}

{\bf Ref.~\cite{Sli90Principles} (see particularly Sec.~8.8 and
  App.~E)}

\medskip

\noindent Nuclear magnetic resonance has been the main tool to study
echoes generated by different time-reversal procedures since the
fifties. After the time-reversal of \textit{individual spin}
precession implemented by Erwin Hahn in the \textbf{spin echo}, one
had to wait until the seventies, when Rhim and Kessemeir achieved the
reversion of the spin-spin interactions. The resulting reversal of the
macroscopic polarization is known as \textbf{Magic Echo} and was
thoroughly studied by Rhim, Pines and Waugh. In the nineties Ernst et
al.  implemented a strategy to address a localized spin excitation and
Levstein et al. realized that such study was optimal for a theoretical
description. They used crystal structures with abundant of interacting
$^{1}$H spins, where the dipolar coupling produces the "diffusion" of
the initial polarization. The time-reversal procedure is then able to
refocus the excitation generating a \textbf{Polarization Echo}. These
last experiments have a direct connection with the concepts discussed
in this review.

\subsubsection{Basic concepts on spin dynamics}

{\bf Refs.~\cite{PLU95Quantum, MBS+97Time, LUP98Attenuation,
    DPL04Spin, ZDL+Quantifying}}

\medskip

\noindent A typical experiment starts with a sample which constitutes
a network with about $10^{23}$ interacting spins in thermal
equilibrium in presence of an external magnetic field $B_{0}$. This
many-spin state is denoted as $|\Psi_{\mathrm{eq}}\rangle$. In this
state every spin has an almost equal probability of being up or down
since the thermal energy, $k_{B}T$, is much higher than any other
relevant energy scale of the system. At time $t=0$ a sequence of
radiofrequency pulses ensures that spin at site $0$th is oriented
along $z$ direction. This is represented by the action of the spin
operator $S_{0}^{+}$ on $|\psi_{\rm eq}\rangle$. In a system with $m+1$
spins, we represent this initially excited state as
\begin{equation}
|\Psi_{0}\rangle=\frac{S_{0}^{+}|\Psi_{\mathrm{eq}}\rangle}{\left\vert
  \langle\Psi_{\mathrm{eq}}|\left.  S_{0}^{-}S_{0}^{+}\right\vert
  \Psi_{\mathrm{eq}}\rangle\right\vert ^{1/2}}=\sum_{r=1}^{2^{m}}%
\frac{e^{\mathrm{i}\phi_{r}}}{2^{m/2}}\left\vert
\uparrow_{0}\right\rangle \otimes\left\vert \beta_{r}\right\rangle
,\label{initial-spin-state}
\end{equation}
where the denominator $\left\vert \langle\Psi_{\mathrm{eq}}|\left.
S_{0}^{-} S_{0}^{+}\right\vert \Psi_{\mathrm{eq}}\rangle\right\vert
^{1/2}$ ensures that the initially excited state has a proper
normalization and $\phi_{r}$ is a random phase that describe a mixture
of states of the form
\begin{equation}
\left\vert \beta_{r}\right\rangle =\left\vert s_{1}\right\rangle
\otimes\left\vert s_{2}\right\rangle \otimes\left\vert
s_{3}\right\rangle \otimes...\otimes\left\vert s_{m}\right\rangle
\ \text{\ with }\left\vert s_{k}\right\rangle \in\left\{ \left\vert
\uparrow\right\rangle ,\left\vert \downarrow\right\rangle \right\}
.\label{mixture-terms}
\end{equation}

The preparation of this state involves the use of a $^{13}$C nucleus
as a \textquotedblleft spy\textquotedblright\ to inject and detect
polarization at the directly bonded $^{1}$H spin (0th). Then, the
system evolves under mutual many-body interaction described by $H_{1}$
for a period $t_{1}$. In the polarization echo experiment, this is
described by effective dipolar interaction Hamiltonian
$H_{1}$, truncated by the Zeeman field,
\begin{equation}
H_{1}\underset{%
%TCIMACRO{\QATOP{\text{Polarization}}{\text{Echo Experiment}}}%
%BeginExpansion
\genfrac{}{}{0pt}{}{\text{Polarization}}{\text{Echo Experiment}}%
%EndExpansion
}{\equiv}%
%TCIMACRO{\dsum \limits_{i,j}}%
%BeginExpansion
{\displaystyle\sum\limits_{i,j}}
%EndExpansion
\left[
  2S_{i}^{z}S_{j}^{z}-\frac{1}{2}(S_{i}^{+}S_{j}^{-}+S_{i}^{-}S_{j}^{+})\right]
.\label{Hdipolar}
\end{equation}
$H_{1}$ contains flip-flop or XY operators of the form
$S_{i}^{+}S_{j}^{-}+S_{i}^{-}S_{j}^{+}$, as well as Ising terms of the
form $S_{i}^{z}S_{j}^{z}$. The dipolar interaction $d_{i,j}$ constant
decay with the third power of the distance between sites $i$ and
$j$. $H_{1}$ produces the spread of the initially localized
excitation. Since total polarization is conserved under $H_{1}$, such
process is commonly known as \textquotedblleft spin
diffusion\textquotedblright.  \ Then, a new pulse sequence rotates all
the spins 90 degrees and the irradiation with r.f. field produces a
further truncation of the dipolar interaction along the rotating
field. The resulting effective Hamiltonian $-H_{2}=-(H_{1}+\Sigma)$
acts for another period $t_{2}$. Observe that the reversal is not
perfect but $\Sigma$ is a small term which is due to truncations
required to form the effective Hamiltonian. This constitutes the
\textquotedblleft backwards\textquotedblright\ evolution period after
which the local polarization at site $0$th is measured. We notice that
$\Sigma$ \ acts as a self-energy operator that may account for
interactions within the same Hilbert space, and thus is just an
Hermitian effective potential.  However, it also may describe the
interaction with a different subsystem that constitutes the
"environment". In the last case $\Sigma=\Delta-\mathrm{i}\Gamma$ would
also have an imaginary (non-Hermitian) component. In this last case,
evolution of observables should be described by two coupled Lindblad
or Keldysh equations for the density matrix requiring a
self-consistent evaluation of dynamics of the system and the
environment. This last situation would make the analysis less
straightforward and might hinder the essential physics. Thus, we first
focus on perturbations that ensure unitary evolution.  An example
discussed by Zangara et al. of this situation is a second frozen spin
chain interacting with the first through an Ising interaction. In
these cases, a polarization echo is formed when $t_{2}=t_{1}$,
i.e. after a total evolution time $2t=t_{2}+t_{1}$. The normalized
amplitude is given by the spin autocorrelation function that accounts
for the observed local polarization,
\begin{equation}
M_{\mathrm{PE}}(t)=\frac{\langle\Psi_{\mathrm{eq}}|S_{0}^{-}(2t)~S_{0}^{+}
  |\Psi_{\mathrm{eq}}\rangle}{\langle\Psi_{\mathrm{eq}}|S_{0}^{-}S_{0}^{+}
  |\Psi_{\mathrm{eq}}\rangle} \,, \label{PolEcho_definition}
\end{equation}
where $S_{0}^{-}(2t)=e^{iH_{1}t/\hbar}e^{-iH_{2}t/\hbar}S_{0}^{-}
e^{iH_{2}t/\hbar}e^{-iH_{1}t/\hbar}\,$\ is the local spin operator,
expressed in the Heisenberg representation respect to the acting
Hamiltonian $H(\tau)=H_{1}\theta(t-\tau)-H_{2}\theta(\tau-t)$. An
equivalent definition of $M_{\mathrm{PE}}(t)$ in terms of
$S_{0}^{z}(2t)~S_{0}^{z}$ \ was also proved useful.

The connection of this experimental observable with the concepts
discussed so far are already hinted but not yet clarified by
Eq.~(\ref{PolEcho_definition}). Thus, we consider that the $m+1$ spins
are arranged in a linear chain or in an odd sized ring where spin-spin
interaction is restricted to XY terms acting on nearest-neighbors. In
this case the Wigner-Jordan spin-fermion mapping allows to describe
the spins system as a gas of non-interacting fermions in a 1-$d$
lattice. In that problem, the initial excitation propagates as a
single spinless fermion. In turn, the observed local polarization can
be written back in terms of the evolution of an initially localized
``spin wave'':
\begin{equation}
|\psi_{0}\rangle=\left\vert \uparrow_{0}\right\rangle
\otimes\left\vert \downarrow_{1}\right\rangle \otimes\left\vert
\downarrow_{2}\right\rangle \otimes\left\vert
\downarrow_{3}\right\rangle \otimes...\otimes\left\vert
\downarrow_{m}\right\rangle \ \text{,}\label{chain-state}
\end{equation}
where
%TCIMACRO{\TEXTsymbol{\vert}}%
%BeginExpansion
$\vert$%
%EndExpansion
$\left\vert \uparrow_{n}\right\rangle $ ($\left\vert \downarrow_{n}
\right\rangle $) means that the $n$-th site is occupied
(unoccupied). Thus, it is clear that the excitation at site $0$-th can
spread along the 1-$d$ chain through the nearest neighbor flip-flop
interaction. This correspondence between many-body dynamics and spin
wave behavior was worked out in detail by Danieli et al. This
procedure was originally employed by Pastawski et al. in the context
of time-reversal experiments to predict the presence of Poincar\'{e}
recurrences (Mesoscopic Echoes) which were clearly observed by
M\'{a}di et al. at the laboratory of Richard Ernst in Zurich. In this
condition, the polarization detected after the time-reversal procedure
results:
\begin{equation}
M_{\mathrm{PE}}(t)\underset{%
%TCIMACRO{\QATOP{1d~\text{system}}{\text{with }XY\text{ interaction.}}}%
%BeginExpansion
\genfrac{}{}{0pt}{}{1d~\text{system}}{\text{with }XY\text{ interaction}}%
%EndExpansion
%\text{. }
}{\equiv}M(t)=\left\vert \langle\psi_{0}|e^{iH_{2}t/\hbar}
e^{-iH_{1}t/\hbar}|\psi_{0}\rangle\right\vert ^{2}.\label{LE-chain}
\end{equation}
Notice that here $|\psi_{0}\rangle$ denotes a single particle wave
function, there are no other operators than the propagators and the
modulus square makes explicit the positivity of
$M_{\mathrm{PE}}(t)$. These features that were not obvious in
Eq.~(\ref{PolEcho_definition}) and make it to agree with the
definition of the Loschmidt echo of Eq.~(\ref{LE_definition}).

Many other effective many-body Halmiltonians can be experimentally
reversed to obtain the revival of different observables that reduce to
a detectable polarization. Thus it is now a common practice to losely
call \textbf{Loschmidt Echo} to the polarization resulting from any of
these possible spin dynamics reversal procedures, regardless the
applicability of the single particle correspondence, with an explicit
mention of the observable used.

\subsubsection{Time-reversal of the dipolar many-spin interaction}

{\bf Refs.~\cite{LUP98Attenuation, UPL98Gaussian, PLU+00nuclear,
    ZCP07Gaussian}}

\medskip

\begin{figure}[ht]
\center
\includegraphics[width=3.5in]{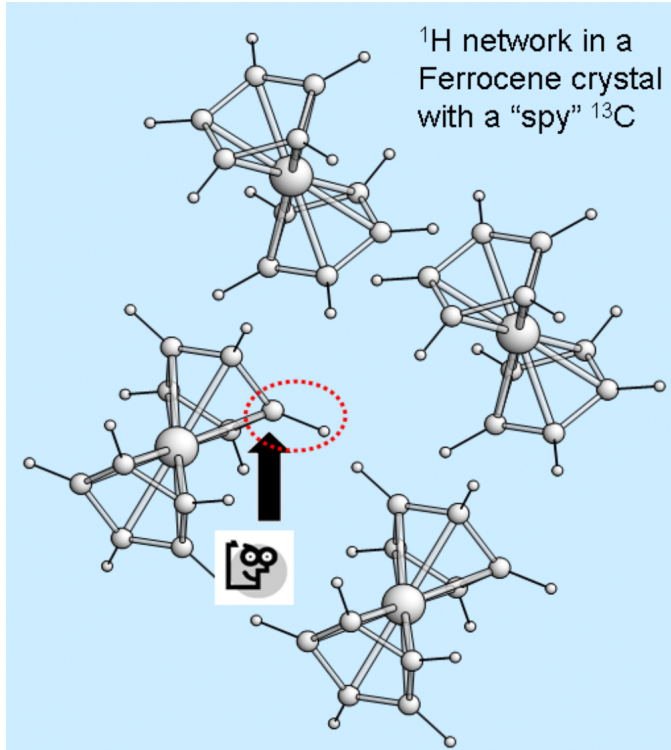}
\caption{Ferrocene crystal showing the $^1$H (small dots), C atoms
  (medium-sized dots) and Fe (large dots). A rare $^{13}$C nucleus
  (usually called a ``spy'' nucleus), acting as a local probe, is
  shown by the arrow. Polarization evolves in the H lattice. Fe atoms
  at the center are not magnetically active, but when replaced by Co
  atoms relaxation of polarization is achieved. Adapted from
  Ref.~\cite{LUP98Attenuation}. Copyright (1998), American Institute
  of Physics.
  %Ferrocene crystal showing the $^{1}$H as small circles. A
  %rare $^{13}$C nucleus (usually called a \textquotedblleft
  %spy\textquotedblright\ nucleus), acting as a local probe, is shown
  %by the arrow. Polarization evolves in the H lattice. Fe atoms at the
  %center are not magnetically active, but in some experiments it is
  %replaced by Co to achieve relaxation.
}
\label{Ferrocene-Xtal}
\end{figure}

\noindent The experiments have been carried out in different molecular
crystals of the cyclopentadienyl (C$_{5}$H$_{5}$) family. There,
$^{1}$H nuclei are arranged in five fold rings with strong intra-ring
dipolar interactions which nevertheless are not constrained to the
molecule but also yield appreciable intermolecular
couplings. Polarization can be injected on those $^{1}$H which are
close to a ``spy'' $^{13}$C nucleus that acts as a local probe used to
inject and eventually detect local polarization as represented in
Fig.~\ref{Ferrocene-Xtal}.
Spreading of the initally localized spin excitation proceeds through a
dipolar interaction which is truncated either in the basis of external
magnetic field (Laboratory frame) or in the rotating frame of a radio
frequency field. This choice is crucial in order to achieve the change
of the sign and the eventual scale down of the effective
Hamiltonian. Thus a spin dynamics with $H_{1}$ in the Laboratory frame
is reversed by a dynamics of $H_{2}$ in the rotating frame or vice
versa. The role of environment might be played by paramagnetic Co(II)
or quadrupolar Mn nuclei that replace the Fe and thus they are absent
for pure ferrocene crystals. Besides, in pure ferrocene truncation in
the rotating frame give rise to non-inverted non-secular terms which,
being supressed by the corresponding Zeeman energy, are at most of the
order of a few percent of the matrix elements of $H_{2}$. This would
yield an Hermitian $\Sigma$. The strength of these terms is inversely
proportional to $B_{1}$, the strength of the r.f. pulse, and thus one
has the possibility to experimentally reduce its importance by
increasing the r.f. power.

\begin{figure}[ht]
\center
\includegraphics[width=4in]{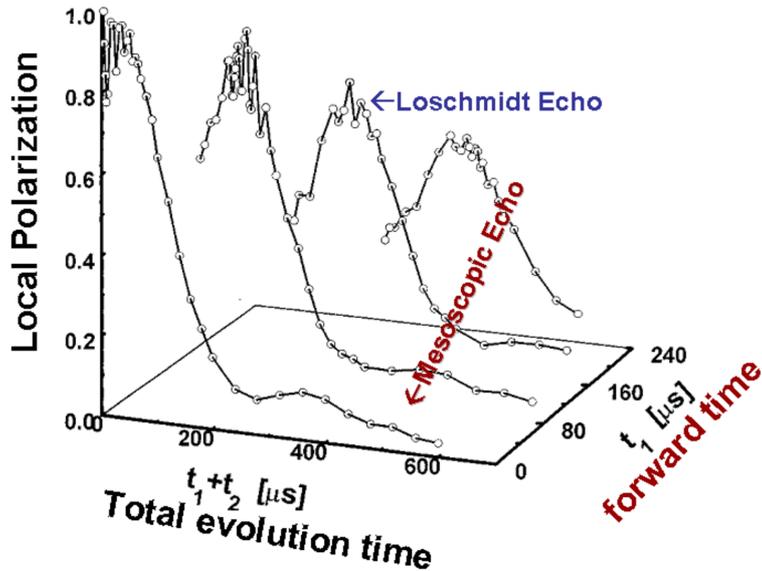}
\caption{Formation and fading away of the Loschmidt echo with reversal
  time $t_{2\text{ \ }}$ elapsed after different periods $t_{1\text{
  }}$of forward evolution. Notice that the tails of the first
  Loschmidt Echo reflect the forward dynamics. At longer times there
  is a first revival asociated with the neighbors and a small revival
  associated with the Mesoscopic Echo resulting from the dynamics
  within the ring. Superimposed with the Loschmidt echo there are high
  frequency oscillations. These are coherent interferences resulting
  from an incomplete transfer between the $^{13}$C, used as a local
  probe, and the nearby $^{1}$H nuclei. Adapted from
  Ref.~\cite{LUP98Attenuation}. Copyright (1998), American Institute
  of Physics.}
\label{LoschmidtEchoFormation}
\end{figure}

The results showing the buildup of the Loschmidt echoes, under $H_{2}$
after different periods of evolution with $H_{1}$, are presented in
Fig.~\ref{LoschmidtEchoFormation}.
The universal features of the Loschmidt echo appear when one studies
the maximum recovered polarization (Loschmidt Echo intensity) as a
function of $t$, the evolution time before reversal. In Ferrocene
crystals, once one substrates the background noise it is clear that
the Loschmidt echo follows a clear Gaussian law,
\begin{equation}
M(t)=\exp\left[-\frac{1}{2}\left(\frac{t}{T_{3}}\right)^{2}\right] \,,
\label{Loschmidt-Gaussian}
\end{equation}
for over two orders of magnitude and serves as a definition of the
characteristic time $T_3$. The same decay law is observed when, using
a specially tailored pulse sequence, the Hamiltonians are scaled down
by a factor $n=1,2,8$ and $16$ respectively. i.e. the characteristic
decay time is increased by the facton $n$. The scaling of $1/T_{3}$
with the Hamiltonian strength is also observed when $H_{1}$ and
$H_{2}$ are scaled down by using different crystal orientations.

\begin{figure}[ht]
\center
\includegraphics[width=4in]{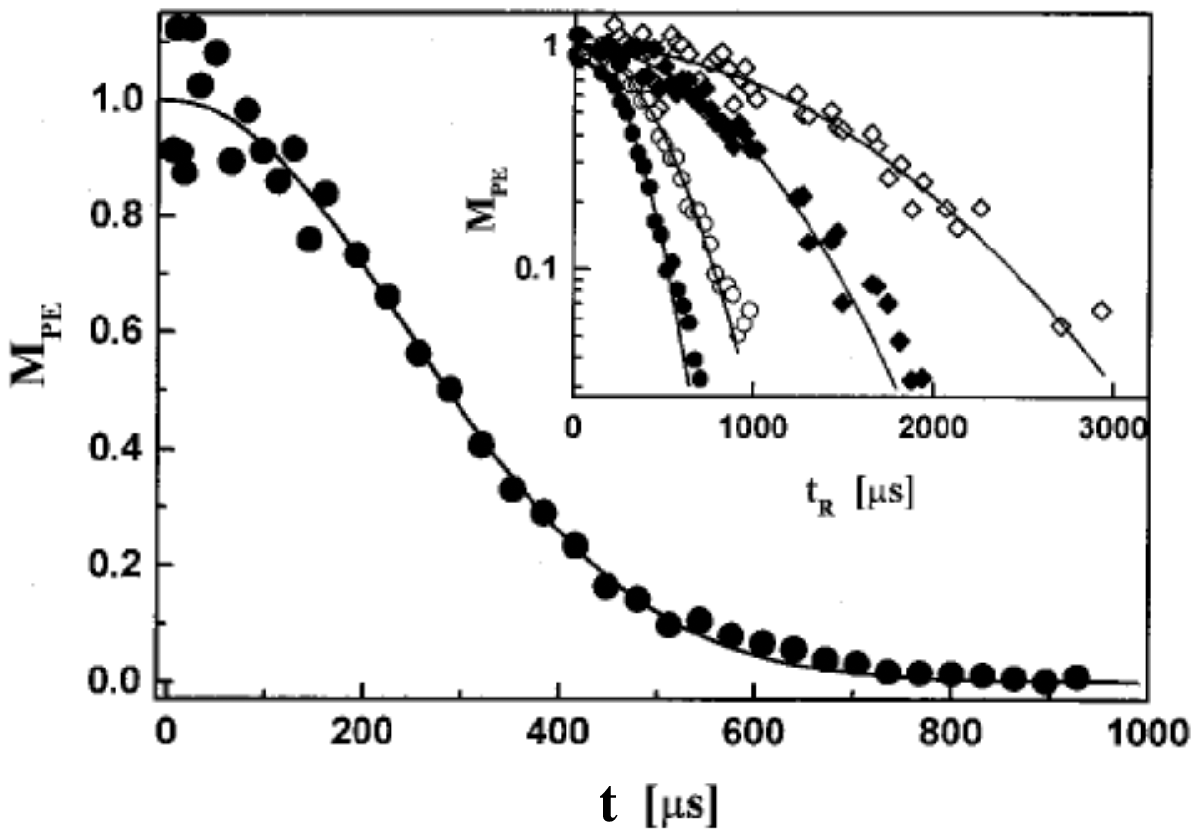}
\caption{Attenuation of the polarization echo in the ferrocene single
  crystal as a function of $t$ for $n=1$ (REPE sequence). The line
  represents a Gaussian fitting. The inset shows the PE attenuation
  for progressively reduced dipolar dynamics, $n=$ 1, 2, 8, and 16. No
  asymptotic regime is reached within the experimental timescale. The
  solid lines are Gaussian fittings. Adapted from
  Ref.~\cite{UPL98Gaussian}.}
\label{Ferrocene-Loschmidt-Scaling}
\end{figure}

In contrast to Ferrocene, in Cobaltoncene crystals, one starts with a
Gaussian decay, and as the dipolar Hamiltonian is scaled down, an
exponential decay of the Loschmidt echo is developed and became stable
with a characteristic time $\tau_{SE}$.
\begin{figure}[ht]
\center
\includegraphics[width=4in]{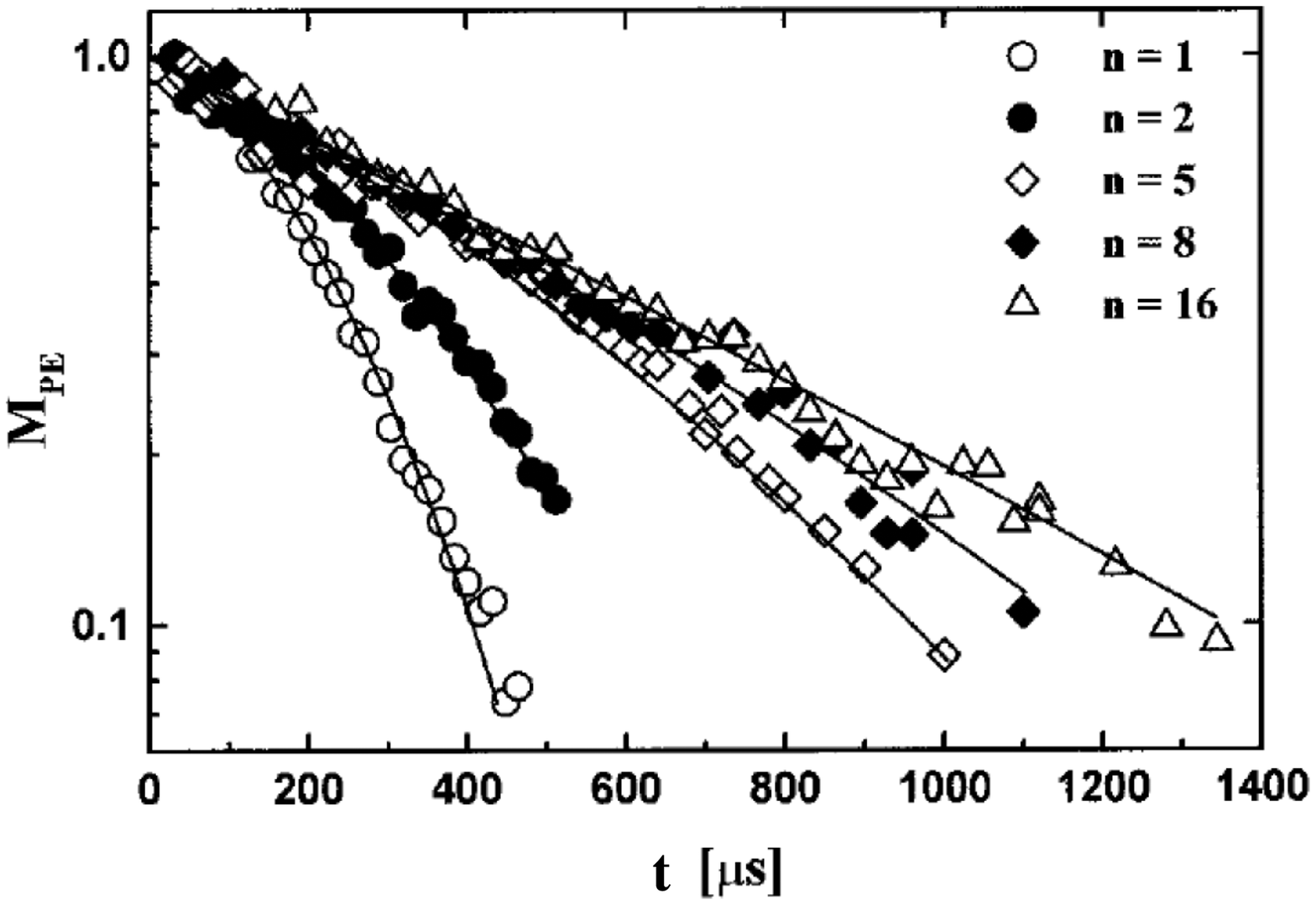}
\caption{Attenuation of the polarization echo in the cobaltocene
  sample as a function of $t$ for $n$ = 1, 2, 5, 8, 16. The
  experimental data show a clear crossover between a dominant Gaussian
  attenuation to an exponential one.  The solid lines represent
  fittings of the whole set of data to
  Eq.~(\ref{PolarizationEchoDecay}). Adapted from
  Ref.~\cite{UPL98Gaussian}.}
\label{Loschmidt-Cobaltocene}
\end{figure}
Thus, the overall behavior of the recovered local polarization is
\begin{equation}
  M(t)=\exp\left[-\frac{1}{2}\left( \frac{t}{nT_{3}} \right)^{2} -
    \frac{t}{\tau_{SE}} \right] \,.
\label{PolarizationEchoDecay}
\end{equation}
One may assign this asymptotic exponential decay rate $1/\tau_{SE}$ to a Fermi
Golden Rule decoherence rate induced by the perturbation induced by the
paramagnetic nature of the Co(II) nuclei which acts as an environment. Further
experiments by Pastawski et al. in crystals free of magnetic impurities
confirmed the Gaussian decay with a rate $1/T_{3}$ that depends only weakly on
the r.f. power and, when truncation terms becomes too small, saturates to an
intrinsic value that scales with the strength of the reverted Hamiltonian.

In summary, these experiments hinted that there is an intrinsic
decoherence rate which is fixed by the inverted Hamiltonian. Thus,
other than the difference between the Gaussian and exponential decay,
one might say that $1/T_{3}$ plays a role similar to the Lyapunov
exponent in the reversal of chaotic one body systems. This surprising
finding suggested that, even in absence of any important perturbation,
in the experimental limit of $m\rightarrow\infty$, even very small
residual terms, in presence of the dynamical instability of a very
complex many-body dynamics, are efficient enough to set the Loschmidt
Echo decay into a perturbation independent decay regime. This plays
the role of an ``intrinsic decoherence'' with a time scale determined
by the inverted Hamiltonian.

It was precisely the hypothesis of an ``intrinsic decoherence rate''
what triggered the theoretical analysis of time-reversal in chaotic
systems. Indeed, while dynamics of one body systems in 1D, described
by Eq.~(\ref{LE-chain}), can not be chaotic, one deems that disorder
and the more complex many-body dynamics present in higher dimensional
systems, have mixing properties which makes them assimilable to
chaotic systems. This observation led to G. Usaj et al. to propose
that quantum chaos contains the underlying physics of a time-reversal
experiment in a many-body systems. It was also argued that
$M_{\mathrm{PE}}(t_{R})$ constitutes an entropy measure.  These ideas
finally boiled down into the model proposed by Jalabert and
Pastawski. While Gaussian decays appears quite naturally as a
consequence of the large number statistics of many-spin states that
are progressively incorporated into the dynamics, the perturbation
independent decay have not found yet a straightforward explanation in
this context.

\subsubsection{Further time-reversal Experiments}

{\bf Refs.~\cite{LCP+04NMR, SPL07Time, RSO+09Effective, SLA+09NMR,
    RCV+11Experimental}}

\medskip

\noindent While time-reversal of different interactions is almost an
unavoidable tool in experimental NMR techniques, there are not so many
studies devoted to grasp at the origins of the (in)efficiency of such
procedures. In particular, time-reversal was implemented using
different procedures and systems that range from 3$d$ and quasi-1$d$
crystals to molecules in oriented liquid crystalline phases. In these
cases, also different initial states and effective Hamiltonians were
studied.

Of particular interest are the double quantum ($DQ$) Hamiltonians,
\begin{equation}
H_{1}\underset{%
%TCIMACRO{\QATOP{\text{Loschmidt}}{\text{Echo for }DQ}}%
%BeginExpansion
\genfrac{}{}{0pt}{}{\text{Loschmidt}}{\text{Echo for }DQ}%
%EndExpansion
\text{ }}{\equiv}H_{DQ}=%
%TCIMACRO{\dsum \limits_{i,j}}%
%BeginExpansion
{\displaystyle\sum\limits_{i,j}}
%EndExpansion
\widetilde{d}_{i,j}\left[  S_{i}^{+}S_{j}^{+}+S_{i}^{-}S_{j}^{-}\right] \,.
\label{HDQ}
\end{equation}
The notation $\widetilde{d}$ for the interaction strength recalls at
it is an effective Hamiltonians built up from dipole-dipole
interaction by suitable pulse sequences whose effect is described with
the help of the average Hamiltonian theory. Since $H_{DQ}$ does no
commute with the polarization described by $\sum_{i}S_{i}^{z},$ it
gives a more mixing as dynamics explores different subspaces of the
Hilbert space inducing the multiple quantum coherences. Besides
$H_{DQ}$, there are high order terms that act as an environment whose
instantaneous decoherence could be described by a Fermi Golden
Rule. One experimental observation is that the larger the number of
subspaces correlated by the interaction (coherences of high order) the
faster the decoherence.

One of the interesting properties of 1$d$ systems, is that the
dynamics induced by a $H_{DQ}$ has a precise correspondence with an XY
chain. Thus, the Loschmidt Echo is given by Eq.~(\ref{LE-chain}). In
real systems, such as Hydroxyapatite (HAp), inter-chain interactions
are significative, quantum Zeno effect induced by a strong dynamics
along the chain prevents the development of coherences beyond second
order and Eq.~(\ref{LE-chain}) remains valid. This explains why the
Loschmidt echo observed by Rufeil-Fiori et al. for the double quantum
Hamiltonian shows a Loschmidt echo with an exponential decay described
by the Fermi golden rule.

\begin{figure}[ht]
\center
\includegraphics[width=3.6in]{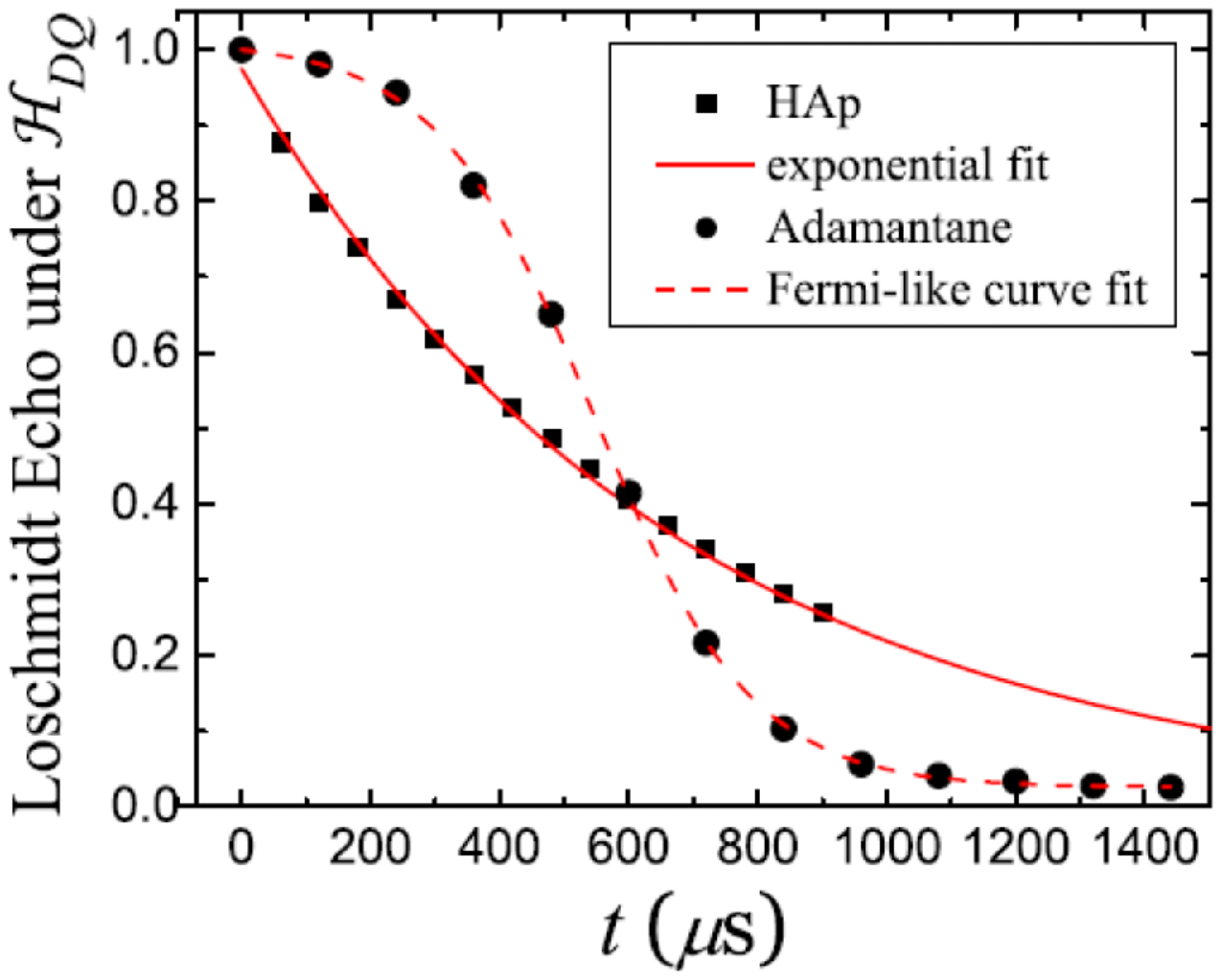}
\caption{Loschmidt Echo experiment based on $H_{DQ}$ and -$H_{DQ}$,
  for hydroxyapatite (HAp) (quasi-1$d$ crystal) and Adamantane, a
  fully 3-$d$ crystal. Here, the source of $\Sigma$ are very small
  high order terms in the effective Hamiltonian.}
\label{DQ-Loschmidt}
\end{figure}

This decay contrasts with results for an Adamantane crystal, a highly
connected 3$d$ system. There, a Fermi-function like decay is
indicative of the two time scales. Each molecule has 16 spins which do
not have direct interactions, thus remain independent for short times
until they interact through neighbor molecules. In this case
decoherence remains weak while intermolecular correlation builds
up. Once neighboring molecules become fully coupled the density of
directly connected states becomes very high and the Fermi-golden-rule
controlled exponential decay takes over.

Spin dynamics in various liquid crystals was studied in series of
papers whose aim was to reduce the number of interacting spins to
those within each molecule. However, the experiences have shown that a
number of residual interaction remain significative and this strongly
compromises the effectiveness of the Loschmidt echo sequences.

\subsection{Microwave billiards}

{\bf Refs.~\cite{SGSS05Fidelity, SSGS05Experimental, HKS08Algebraic,
    BZK+09Probing, KKS+10Microwave, KKS+11Fidelity}}

\medskip

\noindent Microwave-frequency electromagnetic waves in
quasi-two-dimensional cavities are effectively governed by the
Helmholtz equation, which is equivalent to the stationary
Schr{\"o}dinger equation. This equivalence allows one to address the
Loschmidt echo in laboratory experiments with microwave billiards.

Quantities directly measured in microwave experiments are
frequency-dependent scattering matrix elements, $S_{ab}(\nu)$ and
$S'_{ab}(\nu)$, corresponding to the unperturbed and perturbed chaotic
billiard systems, respectively. Here, the indices $a$ and $b$ refer to
the antennae (or scattering channels) involved in the experiment. The
change from the frequency domain to the time domain is achieved by the
Fourier transforms, $\hat{S}_{ab}(t) = \int d\nu \, e^{-2\pi i \nu t}
S_{ab}(\nu)$ and $\hat{S}'_{ab}(t) = \int d\nu \, e^{-2\pi i \nu t}
S'_{ab}(\nu)$, performed over an appropriate frequency window. The
sensitivity of a scattering process to system perturbations are then
quantified by the {\it scattering fidelity amplitude}

\begin{equation}
  f_{ab}(t) = \frac{\langle \hat{S}_{ab}^*(t) \hat{S}'_{ab}(t)
    \rangle}{\sqrt{\langle |\hat{S}_{ab}(t)|^2 \rangle \langle
      |\hat{S}_{ab}(t)|^2 \rangle}} \,,
\label{eq:scat_fid_ampl}
\end{equation}
where the asterisk stands for the complex conjugation, and the angular
brackets denote an ensemble average over different realizations of the
experiment (e.g., different positions of scatterers, antennae,
etc.). The {\it scattering fidelity} itself is a real-valued quantity
defined as $F_{ab}(t) = |f_{ab}(t)|^2$. The numerator in the
right-hand-side of Eq.~(\ref{eq:scat_fid_ampl}) quantifies
correlations between scattering matrix elements of the unperturbed and
perturbed system. The decay of the numerator however is dominated by
the decay of the autocorrelations, so that the denominator is
introduced to compensate for the autocorrelation contributions to the
correlation decay.

\begin{figure}[ht]
\center
\includegraphics[width=4in]{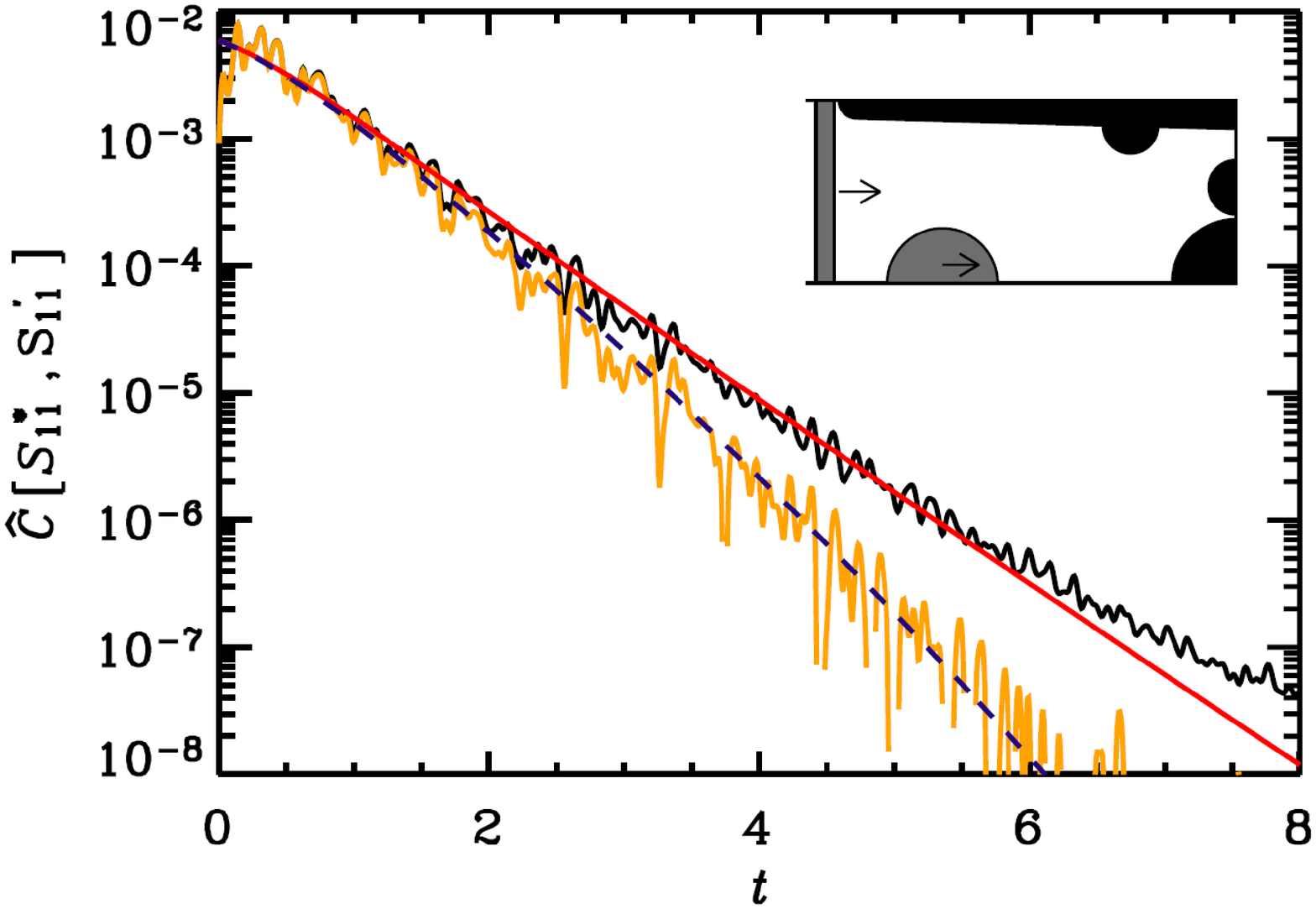}
\caption{Correlation function $\hat{C}[S_{11}^* , S'_{11}] \propto
  \langle \hat{S}_{11}^*(t) \hat{S}'_{11}(t) \rangle$. The
  experimental results for the autocorrelation are shown in black,
  while the correlation of perturbed and unperturbed system are shown
  in gray/orange. The smooth solid curve corresponds to the
  theoretical autocorrelation function, and the dashed curve to the
  product of autocorrelation function and fidelity amplitude.  The
  inset shows the billiard geometry used. Movable parts are marked
  with an arrow. Adapted from
  Ref.~\cite{SSGS05Experimental}. Copyright (2005), American Physical
  Society.}
\label{fig-scat_fidel}
\end{figure}

\begin{figure}[ht]
\center
\includegraphics[width=4in]{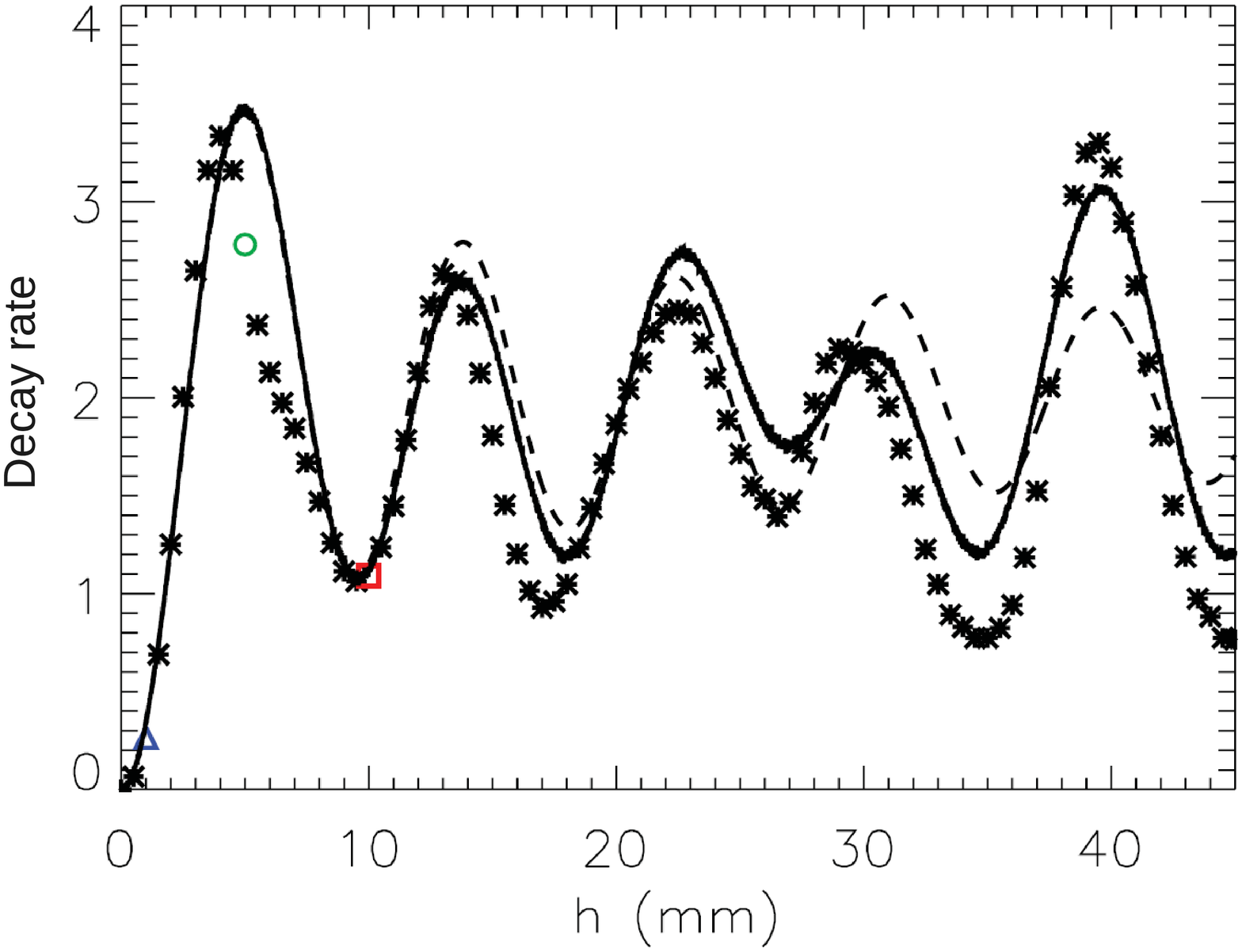}
\caption{Decay rate of the Loschmidt echo in a microwave billiard as a
  function of the displacement $h$ of a small flat part of the
  billiard boundary. The solid (dashed) curve corresponds to the decay
  rate predicted by the semiclassical theory taking (not taking) into
  account classical trajectories multiply scattered within the
  boundary deformation region. Adapted from
  Ref.~\cite{KKS+11Fidelity}. Copyright (2011), American Physical
  Society.}
\label{fig-microwave}
\end{figure}

In chaotic systems and in the case of a weak coupling of the measuring
antennae to the system, the scattering fidelity is known to approach
the Loschmidt echo for a random initial state. This makes microwave
billiards well suited for experimental studies in the field. In
particular, microwave studies provided compelling experimental
evidence for the validity of some of the known decay regimes of the
Loschmidt echo for global and local Hamiltonian perturbations.

\subsection{Elastic waves} 

{\bf Refs.~\cite{LW03Coda, GSW06Scattering}}

\medskip

\noindent Sound waves that travel through a elastic medium are
multiple scattered by its inhomogeneities generating a slowly decaying
wave diffusion.  If no change occurs in the medium over time, the
usually called coda waves are highly repeatable, that is, for
identical excitations the waveforms are indistinguishable. While if
the medium is perturbed, the change in the multiple scattered waves
will result in an observable change in the coda waves.  Lobkis and
Weaver have measured the sensitivity to temperature changes of elastic
coda waves in aluminum alloy blocks.  They used the cross correlation
function between two signals obtained at different temperatures $T_1$
and $T_2$
\begin{equation}
X(\varepsilon)= \frac{\int\rmd t\; S_{T_1}(t)\; S_{T_2}(t(1+\eps))}
   {\sqrt{\int\rmd t\; S_{T_1}^2(t)\; \int\rmd t\; S_{T_2}^2(t(1+\eps))}} \; .
\label{Xeq}\end{equation}
to evaluate the distortion that is defined as $D(t) = -\ln (X_{\rm
  max})$, where the time dependence is given by the ``age'' of the
signal. If formulated as a scattering process, it is shown that $D(t)
= -\ln[f(t)]$, where $f(t)$ is the scattering fidelity that was
introduced to analyze the fidelity decay in microwave billiards. For
sufficiently chaotic dynamics in systems weakly coupled to decay
channels, the scattering fidelity approaches the standard fidelity
amplitude (the Loschmidt echo is the absolute square of the fidelity
amplitude).

\begin{figure}[ht]
\center
\includegraphics[width=4in]{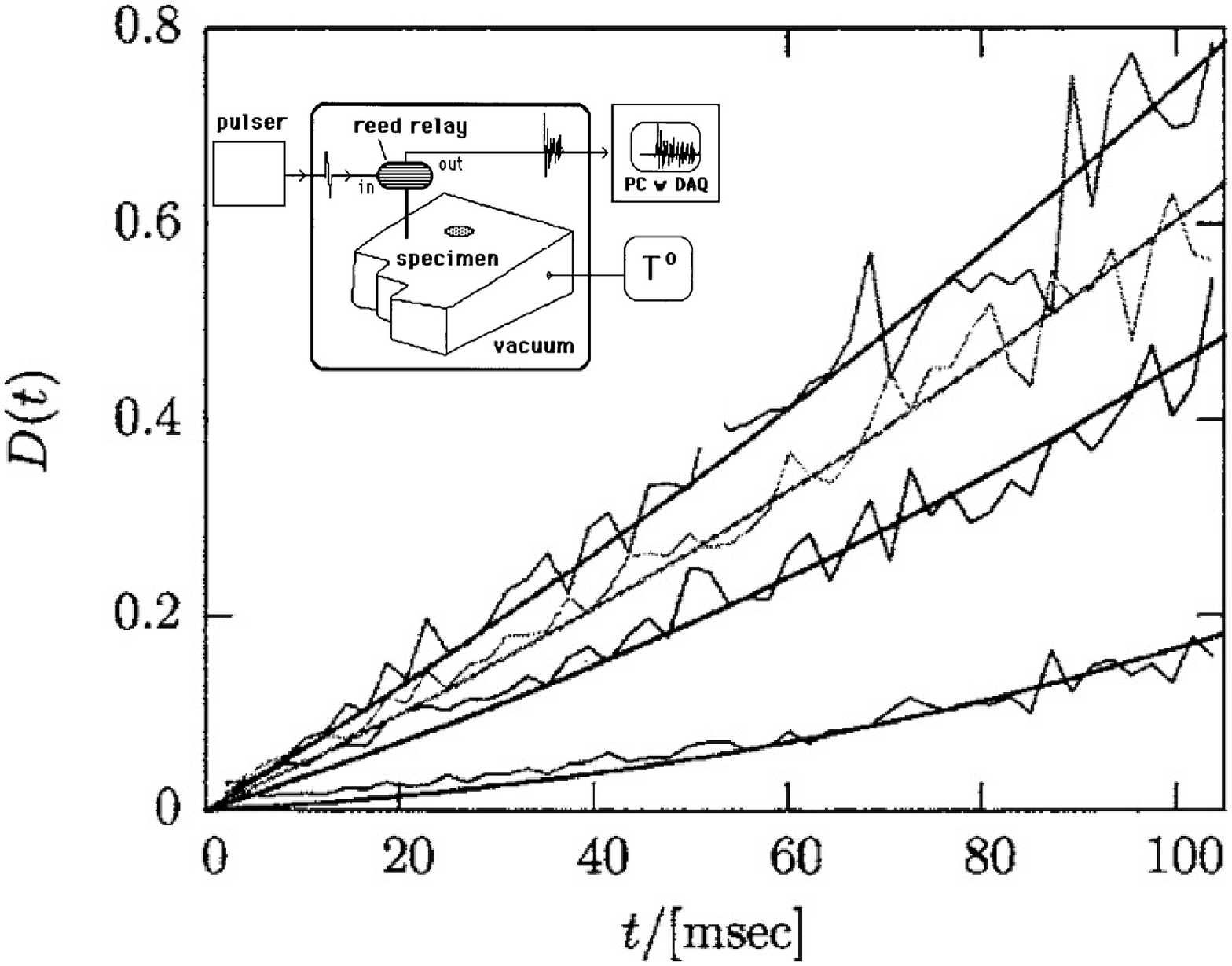}
\caption{The distortion as a function of time for coda waves that
  travel in an aluminum alloy block at different temperatures.  The
  thin lines correspond to measurements in different frequency
  ranges. The thick lines show the corresponding theoretical curves
  obtained using Random matrix theory.  Inset: Laboratory
  configuration. An aluminum specimen, typically with nonparallel
  faces and a defocusing cylindrical hole, is cooled in a vacuum as
  temperature and ultrasonic response are monitored. A reed relay
  isolates the response from the pulser. Adapted from
  Refs.~\cite{LW03Coda} and \cite{GSW06Scattering}. Copyright (2003,
  2006), American Physical Society.}
\label{fig.elastic}
\end{figure}

The results obtained in acoustic signals traveling in an aluminum
blocks were understood using the random matrix predictions for the
standard fidelity (see Fig.~\ref{fig.elastic}). A surprising and
unexpected finding is that the scattering fidelity decay for blocks
with chaotic and regular regular classical dynamics are explained by
the same random matrix expressions.

\subsection{Cold atoms}

{\bf Refs.~\cite{AKD03Echo, AGKD04Revivals, AKGD06Decay,
    WB06Saturation, MGS08Cooling, WTP09Observation, KFOA11Quantum,
    UH11Experimental, DGW12Fidelity}}

\medskip

\noindent The Loschmidt echo, or rather a quantity closely related to
the Loschmidt echo, has been carefully studied in laboratory
experiments with cold atoms trapped inside optical cavities. These
experiments offer an atom-optics realization of the echo in
two-dimensional quantum billiards with underlying chaotic or mixed
classical dynamics.  The basic idea underlying the experiments with
atom-optics billiards is as follows. An effectively two-level atom,
with the internal states denoted by $|1\rangle$ and $|2\rangle$, is
initially prepared in a state $|1\rangle \otimes |\psi\rangle$, where
$|\psi\rangle$ stands for the spatial component of the initial
state. A laboratory observation of the echo involves first exposing
the atom to a sequence of microwave-frequency electromagnetic pulses
and then measuring the probability $P_2$ of finding the atom in the
internal state $|2\rangle$.

In the protocol targeting the Loschmidt echo, $\big| \langle \psi |
e^{i H_2 t / \hbar} e^{-i H_1 t / \hbar} | \psi \rangle \big|^2$, an
atom is first irradiated with a so-called $\pi/2$ pulse, which changes
the atomic state into an equiprobable superposition of $|1\rangle
\otimes |\psi\rangle$ and $|2\rangle \otimes |\psi\rangle$. The pulse
is practically instantaneous and introduces almost no change to the
spatial component of the state. Then, the atom is left to evolve in an
optical trap for a time $t$, during which components $|1\rangle$ and
$|2\rangle$ of the atomic state propagate under Hamiltonians $H_1$ and
$H_2$ respectively. The difference between $H_1$ and $H_2$ originates
from a difference in dipole interaction potentials exerted upon the
states $|1\rangle$ and $|2\rangle$ by the optical trap. Finally,
another $\pi/2$ pulse is applied to the atom and the probability $P_2$
of finding the atom in the internal state $|2\rangle$ is measured;
$P_2$ turns out to be a quantity closely related to the Loschmidt
echo.

Realistic experiments however deal with thermal incoherent mixtures of
initial states, rather than with pure states. As a result, the echo
amplitude, due to an individual state of this thermal mixture,
contributes to $P_2$ with an effectively random phase, smearing out
the echo signal. In order to overcome this difficulty, Andersen,
Davidson, Gr{\"u}nzweig, and Kaplan addressed a new echo measure,
$\big| \langle \psi | e^{i H_2 t / 2\hbar} e^{i H_1 t / 2\hbar} e^{-i
  H_2 t / 2\hbar} e^{-i H_1 t / 2\hbar} | \psi \rangle \big|^2$,
closely related to the Loschmidt echo. In their experiments,
``two-level'' atoms, initially prepared in the state $|1\rangle
\otimes |\psi\rangle$, were successively exposed to (i) a $\pi/2$
pulse, (ii) evolution for a time $t/2$, (iii) a $\pi$ pulse swapping
the population of the internal states $|1\rangle$ and $|2\rangle$,
(iv) another evolution for a time $t/2$, (v) a $\pi/2$ pulse, and
finally (vi) a measurement of $P_2$. Under this protocol, each state
of the thermal mixture contributed to the probability $P_2$ with the
same phase. This allowed for a reliable measurement of the echo even
for ensembles of more than a million of thermally populated states.

More recently, atom interferometry has also been used to investigate
several aspects of time-reversal in quantum kicked rotor systems.  Wu,
Tonyushkin, and Prentiss studied the dynamics of laser-cooled rubidium
atoms subjected to periodically pulsed optical standing waves, as an
atom-optics realization of the kicked rotor. Their experiments have
demonstrated that quantum fidelity of a system, that is chaotic in the
classical limit, can survive strong perturbations over long time
without decay. In a similar setup, Ullah and Hoogerland have performed
an experiment that demonstrated the possibility of using time-reversal
evolution for cooling atomic matter waves. The problem of fidelity
decay in kicked atom-optics systems has recently drawn considerable
interest among theoreticians.

\subsection{Time-reversal mirrors}

{\bf Refs.~\cite{Fin99Time, CJP08Semiclassical, THA+09Sensor}}

\medskip

\noindent Another time-reversal attempt conceptually close to the
Loschmidt echo is that of time-reversal mirrors, developed in the last
twenty years. Such a procedure has been successfully implemented in
various setups where classical waves propagate through a complex media
(going from acoustic to electromagnetic waves). In the time-reversal
mirror protocol an initially localized pulse is recorded by a
collection of receiver-emitter transducers during a time interval
where the wave suffers multiple scattering. The later re-emission of
the time-reversed signal by each transducer during an interval of time
equal to the recording one leads to the refocusing of the signal in
the region of the original excitation.

Two features of this protocol have lead to considerable surprise among
the practitioners:

\begin{itemize}

\item[(i)] even one transducer is enough to obtain a good reproduction
  of the original signal

\item[(ii)] the quality of the refocusing was improved by the
  complexity of the media yielding the multiple scattering of the
  waves.

\end{itemize}

A semiclassical theory of time-reversal focusing can be built up in
terms of propagators and classical trajectories. The two previous
features can be acknowledged by the semiclassical theory.

Time-reversal mirrors and Loschmidt echos differ since the former aims
the refocalization of a wave which is localized in space and time,
while the latter attempt to time-reverse a quantum state. A common
aspect between both protocols is the fact that the Hamiltonian for the
forward and backward evolutions can differ due to modifications of the
environment during the process.

Time-reversal mirrors are not only conceptually important, but also
have very important technological applications as for example brain
therapy, lithotripsy, nondestructive testing and
telecommunications. Recently a sensor of perturbations was proposed
and demonstrated combining the ideas of the Loschmidt echo and time
reversal mirror in classical waves.

\section{Past, present, and future of the Loschmidt echo studies}

Since the early discussions between Boltzmann and Loschmidt on
irreversibility and time-reversal, it has been clearly established
that chaos is the source of the irreversibility in statistical
mechanics. However, when considering the sources of irreversibility in
a quantum system with a few degrees of freedom Quantum Mechanics,
which is the fundamental theory of the microscopic world, does not
allow for chaotic behavior in the sense in which the latter appears in
Classical Mechanics. That is, there is no exponential separation
between states with nearby initial conditions since quantum evolution
is unitary. Trying to understand the origin of irreversibility in
quantum mechanics, Asher Peres proposed in 1984 as an alternative, to
study the stability of quantum motion owing to perturbations in the
Hamiltionian.  In his seminal paper, Peres considered $M(t)$, later
called Loschmidt echo, as a measure of sensibility and reversibility
of quantum evolutions. He aimed to distinguish classically chaotic or
integrable dynamics according to the speed at which the fidelity
decays. Peres reached the conclusion that the long-time behavior of
the fidelity (or saturation, following the terminology of
Sec.~\ref{regimes_ch}) in classically chaotic systems is characterized
by smaller values and smaller fluctuations as compared to the case of
regular dynamics. In view of numerous possibilities of different
behaviors that are allowed in regular systems, such a conclusion does
not always hold.

It is interesting to notice that Peres' article appeared almost at the
same time than two other seminal works studying the relevance of the
underlying classical dynamics for quantum properties. These are the
random-matrix description of the statistical properties of classically
chaotic systems by Oriol Bohigas and collaborators; and the
understanding of the spectral rigidity from a semiclassical analysis
of periodic orbits proposed by Michael Berry. The field of Quantum
Chaos developed building upon these founding ideas and most of the
subsequent works were concerned with the spectral properties of
classically chaotic systems. Comparatively very little work was done
along the Peres' proposal in the decade following it.

Motivated by the puzzles posed by echo experiments in extended dipolar
coupled nuclear spin systems, Rodolfo Jalabert and Horacio Pastawski
studied in 2001 the behavior of the Loschmidt echo for classically
chaotic systems. They found that, depending on the perturbation
strength, the decay of the Loschmidt echo exhibits mainly two
different behaviors. For weak perturbations, the decay is exponential
with the rate that depends on the perturbation strength and that is
given by the width of the local density of states (usually called the
Fermi-golden-rule regime). For stronger perturbations, however, there
is a crossover to a perturbation-independent regime, characterized by
an exponential decay with the rate equal to the average Lyapunov
exponent of the underlying classical system.

The connection of the quantum Loschmidt echo with the classical chaos
generated a great activity in the field. Researches from different
fields, such as quantum chaos, solid state physics, acoustics, and
cold atom physics, have made important contributions towards
understanding various aspects of the Loschmidt echo. During the first
years of the last decade the studies were mainly focused on one-body
aspects of the problem and on the connection between the Loschmidt
echo and decoherence; many interesting experiments were also
performed.

In the last years, the interest has primarily been focused on various
many-body aspects of the Loschmidt echo. For example, there are
studies that consider the decay of the Loschmidt echo as a signature
of a quantum phase transition, or that concentrate on the relation
between the Loschmidt echo and the statistics of the work done by a
quantum critical system when a control parameter is quenched. In view
of recent improvements in experimental many-body system techniques,
the issue of the Loschmidt echo decay has become more concrete.

Some of the more important open questions in the field are:

\begin{itemize}

\item The Lyapunov (perturbation-independent) decay regime has been
  one of the most influential breakthroughs in the theory of the
  Loschmidt echo. However, an experimental observation of the regime
  in simple well-controlled systems is still lacking.

\item The first experimental measure of the Loschmidt echo was done in
  a many-body spin system using NMR techniques. However, theoretically
  little is know about the behavior of the Loschmidt echo in many-body
  systems. Very interesting experiments are now being carried on using
  NMR.

\item The semiclassical theory of the Loschmidt echo has proved to be
  a powerful tool to understand many aspects of its behavior. But this
  approach has some well-recognized difficulties, such as the root
  search problem or the exponentially growing number of classical
  orbits needed for the semiclassical expansion. Recently, a simple
  semiclasical dephasing representation of the Loschmidt echo was
  proposed that does not suffer of the usual problems of the
  semiclassical theories. The range of validity is however unknown.
  
\item There is a vast amount of work characterizing the decay regimes
  of the Loschmidt echo as universal. For chaotic systems perturbed
  with global or local perturbations the behavior of the decay rate of
  the Loschmidt echo is well established (see
  Sec.~\ref{results}). However, recent works in quantum maps that are
  perturbed in all phase space, have reported non-universal
  oscillatory behavior in the decay rate of the Loschmidt echo as a
  function of the perturbation strength. Deviations from the
  perturbation independence are found usually in the form of
  oscillations around the Lyapunov exponent. Moreover, there are cases
  where deviations are considerably large rendering the Lyapunov
  regime non-existing.  The fully understanding of this non-universal
  behavior, that was also observed in a Josephson flux quit, is still
  lacking \cite{GW11Loschmidt, PD07Fidelity}.
 
\item The semiclassical theory of the Lyapunov decay of the Loschmidt
  echo is based on highly localized initial states and the diagonal
  approximation of Eq. (3) for $M(t)$. But recent results using the
  semiclassical dephasing representation have shown a Lyapunov regime
  in the mean value of the fidelity amplitude $\langle m(t) \rangle$
  \cite{GVW11Semiclassical, ZOdA11Initial}.
 
\item Systems with regular or fully chaotic dynamics are rather
  exceptional in nature. A generic system has mixed phase space
  consisting of integrable islands immersed in a chaotic sea. Few
  general results are known about the Loschmidt echo in this scenario.

\end{itemize}

%%%%%%%%%%%%%%%%%%%%%%%%%%%%%%%%%%%%%%%%%%%%%%%%%%%%%%%%%%%%%%%%%%%%%%%%%%%%

%\bibliographystyle{wmaainf}
%\bibliography{../echo,../mystuff,../numerics}

\section{Recommended reading}

\begin{itemize}

\item R.~A.~Jalabert and H.~M.~Pastawski. ``The semiclassical tool in
  complex physical systems: Mesoscopics and decoherence''. {\it
    Advances in Solid State Physics}, Vol. 41, p. 483, 2001.

\item H.~M.~Pastawski, G.~Usaj, and P.~R.~Levstein. ``Quantum Chaos:
  an answer to the Boltzmann-Loschmidt controversy? An experimental
  approach''. \textit{Contemporary Problems of Condensed Matter
    Physics}, pp. 223--258, S.~J.~Vlaev, L.~M.~Gaggero~Sager, and
  V.~V.~Dvoeglazov, Eds., NOVA Scientific Publishers, New York,
  2001. URL: \url{http://web.utk.edu/~pasi/lectures.html}

\item T.~Gorin, T.~Prosen, T.~H.~Seligman, and M.~Znidaric. ``Dynamics
  of Loschmidt echos and fielity decay''. {\it Physics Reports},
  Vol. 435, pp. 33--156, 2006.

\item Ph.~Jacquod and C.~Petitjean. ``Decoherence, entanglement and
  irreversibility in quantum dynamical systems with few degrees of
  freedom''. {\it Advances in Physics}, Vol. 58, pp. 67--196, 2009.

\item A.~Peres. ``Quantum Theory: Concepts and Methods (Fundamental
  Theories of Physcis)''. Springer, 1995.

\item A.~M.~Ozorio~de~Almeida. ``Hamiltonian Systems, Chaos and
  Quantization''. Cambridge Press, Cambridge 1988.

\item F.~Haake. ``Quantum Signatures of Chaos''. Springer-Verlag,
  Berlin-Heidelberg, 2001.

\item H.-J.~St\"ockmann. ``Quantum Chaos: An Introduction''.
  Cambridge University Press, 1999.

\end{itemize}

\section{Internal references}

\begin{itemize}

\item Y.~Fyodorov; {\it Random Matrix Theory}, Scholarpedia
  6(3):9886(2011).

\item M.~Gutzwiller; {\it Quantum chaos}, Scholarpedia
  2(12):3146(2007).

\item M.~Raizen and D.~A.~Steck; {\it Cold atom experiments in quantum
  chaos}, Scholarpedia 6(11):10468(2011).

\item H.-J.~St{\"o}ckmann; {\it Microwave billiards and quantum chaos},
  Scholarpedia 5(10):10243(2010).

\end{itemize}

\section{External links}

\begin{itemize}

\item The Loschmidt Echo homepage

%\item Fink's group

\end{itemize}

\end{document}